\documentclass[
aps,
prl,
10pt,
twocolumn,
oneside,
preprintnumbers,
amsmath,
amssymb,
flushbottom,
nofootinbib,
superscriptaddress,
floatfix]{revtex4-2}

\usepackage[T1]{fontenc}
\usepackage[utf8]{inputenc}

\usepackage{physics}
\usepackage{xspace}
\usepackage{slashed}
\usepackage{braket}
\usepackage{leftindex}
\usepackage{graphicx}  
\usepackage{bm}  
\usepackage{booktabs}
\usepackage[normalem]{ulem}
\usepackage{colortbl}
\usepackage[svgnames,dvipsnames,table,x11names]{xcolor}
\usepackage{comment}
\usepackage{enumitem}
\usepackage{mathrsfs}
\usepackage{amsmath,amssymb,amsthm}
\usepackage{mathtools}
\usepackage{rotating}
\usepackage{nicematrix}
\usepackage{hyperref}

\newcommand{\Op}{\mathcal{O}}
\newcommand{\M}{\mathcal{M}}
\newcommand{\wb}{\bar{w}}
\newcommand{\MSbar}{\overline{\rm MS}}
\newcommand{\Eq}[1]{Eq.~\eqref{#1}}
\newcommand{\Fig}[1]{Fig.~\ref{#1}}
\newcommand{\agl}[2]{\langle #1 #2 \rangle}
\newcommand{\sqr}[2]{\lbrack #1 #2 \rbrack}
\newcommand{\ccol}{\cellcolor{gray!30}}
\newcommand{\new}{\cellcolor{blue!20}}

\allowdisplaybreaks

\hypersetup{
    pdfencoding=unicode,
	colorlinks=true,
	urlcolor=blue,
    linkcolor=blue,
    citecolor=blue,
    pdftitle={All-Order Helicity Selection Rules in Effective Field Theories}
}

\makeatletter\g@addto@macro\bfseries{\boldmath}\makeatother

\begin{document}

\title{All-Order Helicity Selection Rules in Effective Field Theories}

\author{Luigi~C.~Bresciani}
\email{luigicarlo.bresciani@phd.unipd.it}
\affiliation{Dipartimento di Fisica e Astronomia ``G.~Galilei'', Università degli Studi di Padova, Via F.~Marzolo 8,
35131 Padova, Italy}
\affiliation{Istituto Nazionale di Fisica Nucleare, Sezione di Padova, Via F.~Marzolo 8, 35131 Padova, Italy}

\author{Nud\v zeim~Selimovi\'c}
\email{nudzeim.selimovic@pd.infn.it}
\affiliation{Dipartimento di Fisica e Astronomia ``G.~Galilei'', Università degli Studi di Padova, Via F.~Marzolo 8,
35131 Padova, Italy}
\affiliation{Istituto Nazionale di Fisica Nucleare, Sezione di Padova, Via F.~Marzolo 8, 35131 Padova, Italy}

\begin{abstract}
Using on-shell methods, we derive an all-order non-renormalization theorem for general four-dimensional effective field theories, relying on Poincaré invariance and unitarity. Each operator $\mathcal O$ is assigned the weights $(w,\bar w)=(\ell-h,\ell+h)$, where $\ell$ and $h$ are the number of particles and total helicity of its minimal configuration. At $L$ loops, the mixing $\mathcal O_j\to\mathcal O_i$ vanishes when $\ell_j-\ell_i=L-1$ and either $w_i<w_j-2(L-1)$ or $\bar w_i<\bar w_j-2(L-1)$, in the absence of non-holomorphic Yukawa couplings. These zeros follow entirely from tree-level data and are  scheme independent under finite renormalizations that preserve the operator-length selection rule. For operators of dimensions five through eight, we identify broad classes of previously unknown zeros at higher loop orders, with particular emphasis on two loops, where the theorem provides direct checks on current calculations, as in the Standard Model Effective Field Theory.
\end{abstract}

\maketitle

\pdfbookmark[1]{Introduction}{Introduction}
\paragraph{\textbf{Introduction.}}

Effective field theories (EFTs) recast the search for physics beyond the Standard Model as the measurement of the coefficients of higher-dimensional operators.
Their renormalization-group running, governed by the anomalous-dimension matrix, connects the high scale where these coefficients are generated to the low scales where they are probed experimentally.

Although 
naive dimensional analysis
permits many operator mixings, anomalous-dimension matrices are often remarkably sparse. 
At one loop, these unexpected zeros were traced back to several complementary structures. 
From an on-shell perspective, unitarity cuts uncovered the selection rules based on helicity \cite{Cheung:2015aba} and angular momentum \cite{Jiang:2020rwz}. 
Additionally, zeros based on color \cite{Bern:2020ikv} and on the operator length \cite{Bern:2019wie} apply beyond one loop, and the latter have been generalized to multiple operator insertions \cite{Cao:2023adc}.
Moreover, the helicity selection rules accounted for the holomorphic structure first observed in the anomalous dimensions of the Standard Model EFT \cite{Jenkins:2013zja,Jenkins:2013wua,Alonso:2013hga,Alonso:2014rga}, which could also be understood through supersymmetric embeddings \cite{Elias-Miro:2014eia}. Complementarily, positivity provides an independent source of constraints on operator mixing \cite{Chala:2023jyx,Chala:2023xjy,Liao:2025npz}. 

Whether comparable selection rules persist at higher loops is less clear and increasingly important in light of
the improved precision, especially at two loops, relevant to new-physics searches 
\cite{Allwicher:2023aql,Stefanek:2024kds,Haisch:2024wnw,Olgoso:2025jot,Mantani:2026fao,Born:2026tgm,Haisch:2026eyq}
as well as
recent calculations in the Standard Model EFT, Low-Energy EFT, and general EFTs
\cite{Born:2026xkr,Ibarra:2024tpt,Zhang:2025ywe,DiNoi:2025tka,Aebischer:2022anv,Born:2024mgz,DiNoi:2024ajj,Duhr:2025zqw,DiNoi:2025arz,Haisch:2025lvd,Haisch:2025vqj,Duhr:2025yor,Banik:2025wpi,Naterop:2024cfx,Naterop:2025lzc,Naterop:2025cwg,Aebischer:2025hsx,Jenkins:2023bls,EliasMiro:2021jgu,Guedes:2025sax,Aebischer:2026zxv,Fonseca:2025zjb,Fonseca:2025cls,Misiak:2025xzq,Aebischer:2025zxg,Aebischer:2025ddl,Henriksson:2025hwi,Henriksson:2025vyi}.
At higher orders, loop-level contributions weaken the tree-level selection rules, while renormalization-scheme choices can obscure which zeros are physically meaningful, as anomalous dimensions acquire a dependence on the renormalization scheme starting at two loops. 

By employing on-shell methods,
already effectively used to organize and compute anomalous dimensions \cite{Caron-Huot:2016cwu,EliasMiro:2020tdv,Baratella:2020lzz,Jiang:2020mhe,Bern:2020ikv,Baratella:2020dvw,AccettulliHuber:2021uoa,EliasMiro:2021jgu,Baratella:2022nog,Machado:2022ozb,Bresciani:2023jsu,Bresciani:2024shu,Aebischer:2025zxg,Aebischer:2025ddl,Wu:2025qto,Baratella:2021guc,Shu:2021qlr},
we show that a substantial structure nevertheless survives.

To formulate these results precisely, we work in a four-dimensional non-Abelian gauge theory in which massless particles of spin $0$ (scalars $\phi$), $\frac{1}{2}$ (chiral fermions $\psi_\alpha,\bar\psi_{\dot\alpha}$), and $1$ (vectors with field strengths decomposed into self-dual and anti-self-dual components $F_{\alpha\beta},\bar F_{\dot\alpha\dot\beta}$) interact through marginal and irrelevant operators.
The higher-dimensional operators in $\mathcal L^{(d)} = \sum_i c_i \, \mathcal O_i$ have a common mass dimension $d > 4$ and constitute an irreducible physical basis.
Renormalization mixes the operators and the associated coefficients $c_i$:
\begin{equation}
    \mu \frac{\dd c_i}{\dd \mu} = \textstyle\sum_j \gamma_{ij}\,c_j\,,
\end{equation}
where $\gamma_{ij} = \sum_{L\ge 1}\gamma_{ij}^{(L)}$ is the ultraviolet (UV) anomalous-dimension matrix, 
expanded at the relevant loop order $L$.

Two quantities, $w$ and $\wb$,
organize everything that follows.
For an on-shell configuration $X$ of $n$ particles with helicities $h_a$, the \textit{holomorphic} and \textit{anti-holomorphic
weights} \cite{Cheung:2015aba} are defined as
\begin{equation}
w(X)=n-\textstyle\sum_a h_a \,,
\qquad
\wb(X)=n+\textstyle\sum_a h_a \,,
\label{eq:weights}
\end{equation}
so that $F,\psi,\phi,\bar\psi,\bar F$ carry $(w,\wb)=(0,2)$, $(\tfrac12,\tfrac32)$, $(1,1)$, $(\tfrac32,\tfrac12)$, $(2,0)$, respectively. 
We define the weights $(w_i,\wb_i)$ associated with an operator $\mathcal O_i$ as those of the leading
minimal configuration $X_i$ produced by a single insertion of $\Op_i$. 
Any tree form factor containing that insertion has $w\ge w_i$ and $\wb\ge\wb_i$ \cite{Cheung:2015aba}: attaching renormalizable vertices can only raise the weight, since every renormalizable tree amplitude
carries $w,\wb \ge 2$, while each internal line costs two units.
The \textit{length} $\ell_i$ of an operator \cite{Bern:2019wie} is defined as the minimal number of particles created by $\mathcal O_i$: $\ell_i = \frac{1}{2}(w_i + \wb_i)$.

As shown below, when the source operator is $L-1$ units longer than the target, helicity selection rules yield zeros in the $L$-loop mixing that are fixed by tree-level data alone. Furthermore, these zeros are scheme independent under finite renormalizations that preserve the operator-length selection rule \cite{Bern:2019wie} and constitute our central result. 

When specialized to two loops, this result constrains the mixing of operators into those one unit shorter.  In addition, for $\ell_i\ge\ell_j$, and conditional on an assumption about complex-collinear factorization, we construct a scheme---which we name the \textit{holomorphic scheme}---in which the mixing vanishes if $w_i<w_j-4$ or $\wb_i<\wb_j-4$.
In a generic scheme, this mixing is generally non-zero but carries no independent two-loop information, as it is determined entirely by one-loop data. This distinction identifies which entries contain genuinely new two-loop information and which are fixed by lower-order data. \smallskip

\pdfbookmark[1]{Anomalous dimensions from unitarity cuts}{Anomalous dimensions from unitarity cuts}
\paragraph{\textbf{Anomalous dimensions from unitarity cuts.}}

To extract the anomalous dimensions, we use the dispersive, on-shell approach of Ref.~\cite{Caron-Huot:2016cwu}. 
Its main result,
\begin{equation}
    e^{-i\pi D} F_i^* = S F_i^*\,,
    \label{eq:chw}
\end{equation}
with $F_i(X;q) = \mel{X}{\mathcal O_i(q)}{0}$ the form factor of the operator $\mathcal O_i(q)$ injecting off-shell momentum $q$, relates the dilatation operator $D=\sum_a p_a^\mu \, \partial/\partial p_a^\mu$, which acts as $-\mu \, \partial_\mu$ on massless form factors in dimensional regularization, to the phase of the scattering matrix $S$. 
Form factors obey the Callan--Symanzik equation \cite{Callan:1970yg,Symanzik:1970rt},
and their dependence on the renormalization scale then encodes the UV mixing of operator insertions, together with $\beta$-function and infrared (IR) contributions.

In this approach, the weights are tracked across the unitarity cuts that build the form factors: sewing the corners $c$ across $k$ cut legs gives
\begin{equation}
\label{eq:sewing}
    w = \textstyle\sum_c w_c - 2k\,,
    \qquad
    \wb = \textstyle\sum_c \wb_c - 2k\,,
\end{equation}
since each cut leg is counted twice in $\sum_c n_c$ while being internal to the sewn object, and the helicities of its two ends, being opposite, cancel in $\sum_a h_a$.

We consider and review the one- and two-loop expansions of \Eq{eq:chw}, which make explicit the ingredients underlying the theorem. Starting at one loop, one finds~\cite{Caron-Huot:2016cwu}
\begin{equation}
    \Delta\gamma^{(1)}_{i j}F^{(0)}_i(X_i) = -\tfrac{1}{\pi}\big[\mathcal M F_j\big]^{(1)}(X_i)\,,
\end{equation}
where $\Delta\gamma_{i j}^{(L)} = \gamma_{i j}^{(L)} - \gamma_{ij}^{(L),\mathrm{IR}}$ and $\mathcal M$ is the renormalizable amplitude entering the two-particle unitarity cut $[\mathcal M F_j]^{(1)} = \mathcal M^{(0)} \otimes_2  F_j^{(0)}$, 
with $\otimes_n$ denoting the integration over the $n$-body phase space of the intermediate states in the product.
$\gamma_{ij}^{(L),\mathrm{IR}}$ collects the IR anomalous dimensions, generated by the soft and collinear singularities of the external states \cite{Sterman:2002qn,Becher:2009cu,Chiu:2009mg}. They dress lower-loop form factors of the same operator on the same external state without changing the particle content or helicities. 
The entry $\gamma^{(L),\text{IR}}_{ij}$ can therefore be non-zero only if a lower-loop form factor of $\mathcal O_j$ is non-zero on $X_i$ \cite{Catani:1998bh}. In particular, $\Delta\gamma_{ij}^{(L)} = \gamma_{ij}^{(L)}$ if $F^{(L')}_j(X_i)=0$ for all $L'<L$, which holds for $\ell_i \leq \ell_j -\max(1,L-1)$ \cite{Bern:2019wie}.
Moreover, renormalizable tree amplitudes with at least four legs carry $w,\wb \ge 4$ in the absence of non-holomorphic Yukawa couplings, otherwise $w,\wb \ge 2$ \cite{Cheung:2015aba}.\footnote{A Yukawa coupling $y\phi\psi^2$ is holomorphic, involving only $\phi$ and not $\bar\phi$, while a companion coupling $y'\bar\phi\psi^2$ makes the sector non-holomorphic as it allows the exceptional amplitudes $\mathcal M^{(0)}(\psi^+\psi^+\psi^+\psi^+)$ and $\mathcal M^{(0)}(\psi^+\psi^+\psi^+\psi^+ F^+ \dots F^+)$ at $w=2$ to be non-zero \cite{Cheung:2015aba}. 
An example is provided by the Standard Model, where Higgs-doublet exchange generates an exceptional amplitude proportional to the product of up-type and down-type Yukawa couplings.}
Adding the further information that any tree form factor of $\mathcal O_j$ has $w \ge w_j$ and $\wb \ge \wb_j$, we find from Eq.~\eqref{eq:sewing} that the above two-particle cut $\mathcal M^{(0)} \otimes_2  F_j^{(0)}$ must carry $w\ge w_j$ and $\wb\ge \wb_j$ ($w\ge w_j-2$ and $\wb\ge \wb_j-2$ if non-holomorphic Yukawa couplings are present).
This is precisely the content of the helicity selection rule of Ref.~\cite{Cheung:2015aba}:
\begin{equation}
    \gamma_{ij}^{(1)} = 0 \qquad \text{if} \qquad w_i < w_j \ \ \lor \ \ \wb_i < \wb_j\,,
    \label{eq:1lhsr}
\end{equation}
absent non-holomorphic Yukawa couplings, which relax the bound by two units.
Moreover, $\gamma_{ij}^{(1)} = 0$ if $\ell_i < \ell_j$, since then no one-loop diagram exists or does not contain a scaleless integral \cite{Bern:2019wie}.

\begin{figure}[t]
    \centering
    \includegraphics[width=\linewidth,page=1]{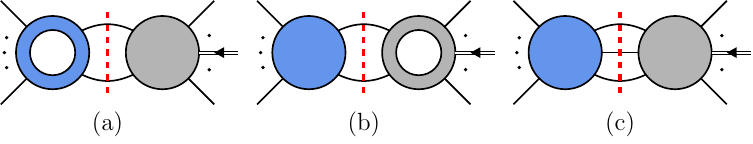}
    \caption{Unitarity cuts relevant for the extraction of anomalous dimensions from two-loop form factors. Higher-dimensional operator insertions are indicated by gray blobs. The blobs with a hole denote one-loop form factors or amplitudes.}
    \label{fig:cuts}
\end{figure}

The two-loop anomalous-dimension matrix can be extracted from the master formula \cite{Bern:2020ikv}
\begin{multline}
\label{eq:master}
\Delta\gamma^{(2)}_{i j}F^{(0)}_i(X_i)
=-\tfrac{1}{\pi}\big[\Re \mathcal M \Re F_j\big]^{(2)}(X_i)\\-\textstyle\sum_k \big[\Delta\gamma_{kj}^{(1)}+\delta_{kj}\beta^{(1)}_g\partial_g\big]\Re F^{(1)}_k(X_i) \,,
\end{multline}
where $\beta^{(L)}_g$ is the $L$-loop $\beta$-function of the marginal couplings, collectively denoted as $g$.
Three unitarity cuts contribute to \Eq{eq:master},
\begin{equation}
\big[\M F_j\big]^{(2)}=
\underbrace{\mathcal M^{(1)}\! \otimes_2 \!F^{(0)}_j}_{\rm (a)}
+\underbrace{\mathcal M^{(0)}\! \otimes_2 \!F^{(1)}_j}_{\rm (b)}
+\underbrace{\M^{(0)}\! \otimes_3 \! F^{(0)}_j}_{\rm (c)} \,,
\label{eq:threecuts}
\end{equation}
and are shown in \Fig{fig:cuts}.

The one-loop logic survives or fails cut by cut.
The genuinely new feature at two loops is that cuts (a) and (b) contain one-loop objects---the renormalized amplitude $\mathcal M^{(1)}$ and the renormalized form factor $F^{(1)}_j$---whose finite rational parts are not determined by four-dimensional unitarity cuts.
It is precisely through such rational terms that one-loop amplitudes evade the tree-level helicity selection rules: for instance, the all-plus vector amplitude is non-zero, is UV and IR finite, and carries the minimal weight $w=0$.
Sewing it into cut (a) through \Eq{eq:sewing} gives $w_i \ge 0 + w_j - 4 = w_j - 4$,
so the one-loop bound can relax by as much as four units.
Whether the two-loop selection rule survives therefore depends entirely on controlling the rational parts of these one-loop objects.
Accordingly, we separate the two-loop zeros into those requiring no rational-part information, which are scheme independent, and those sensitive to it, which are scheme dependent. \smallskip

\pdfbookmark[1]{Scheme-independent zeros}{Scheme-independent two-loop zeros}
\paragraph{\textbf{Scheme-independent zeros.}}

We first establish the selection rule at two loops, where the relevant mixing is fixed by tree-level data alone and is therefore scheme independent. 
Its extension to arbitrary loop order then follows from the same length and helicity counting, as we discuss below.

At two loops, the length selection rule of Ref.~\cite{Bern:2019wie} already forbids a large class of entries: the form-factor side of any cut of $\mathcal O_j$ appearing in Eq.~\eqref{eq:master} needs at least $\ell_j$ legs, so an operator $\mathcal O_j$ cannot renormalize a shorter operator $\mathcal O_i$ at two loops when $\ell_j - \ell_i \ge 2$, since no such diagram can avoid a scaleless integral. Thus, $\gamma^{(2)}_{ij} = 0$ if $\ell_i < \ell_j - 1$.

At the boundary $\ell_i = \ell_j - 1$, since $\gamma_{ij}^{(1)} = 0$, the operator mixing first occurs at two loops, and leg counting collapses the master formula in \Eq{eq:master} onto the single cut (c). 
Indeed, the $\beta$-function term $\delta_{kj}\beta^{(1)}_g\partial_g\Re F^{(1)}_k$ is proportional to a form factor of $\Op_j$ itself, which is non-zero only on an external state with at least $\ell_j$ legs (configurations with fewer legs are forbidden by the particle content at tree level and force scaleless integrals at one loop \cite{Bern:2019wie}), whereas $X_i$ carries only $\ell_i=\ell_j-1$.
The IR term vanishes because it dresses $F^{(0)}_j(X_i)$ and $F^{(1)}_j(X_i)$, which are zero for $\ell_i < \ell_j$.
The iteration term $\Delta\gamma^{(1)}_{kj}\Re F^{(1)}_k(X_i)$ vanishes by the same counting applied twice: $\Delta\gamma^{(1)}_{kj}\neq0$ requires $\ell_k\ge\ell_j$, while $F^{(1)}_k(X_i)\neq0$ requires $\ell_i\ge\ell_k$, so together $\ell_i\ge\ell_j$.
Cuts (a) and (b) are likewise absent: a non-vanishing two-particle cut requires at least four legs on the amplitude side, while tree and one-loop form factors of $\mathcal O_j$ require at least $\ell_j$ legs, since lower-multiplicity one-loop diagrams contain scaleless tadpoles or bubbles \cite{Bern:2019wie,Bern:2020ikv,EliasMiro:2020tdv,Caron-Huot:2016cwu,EliasMiro:2021jgu}. Thus, they yield at least $4+\ell_j-4=\ell_j>\ell_i$ external legs and Eq.~\eqref{eq:master} reduces to
\begin{equation}
    \gamma_{ij}^{(2)} F_{i}^{(0)}(X_i) = -\tfrac{1}{\pi} \big(\M^{(0)}\! \otimes_3 \! F^{(0)}_j\big)(X_i)\,,
\end{equation}
where $\M^{(0)}$ is necessarily a five-point amplitude.
Therefore, the tree-level helicity selection rule promotes to two loops exactly as at one loop. With three sewn legs, \Eq{eq:sewing} gives $w_i \ge 4 + w_j - 6$, so that
\begin{equation}
\label{eq:dl1}
    \gamma^{(2)}_{ij} = 0 \ \ \text{if} \ \ \ell_i = \ell_j - 1 \ \land \  \big(w_i < w_j - 2  \ \lor  \ \wb_i < \wb_j - 2\big) \,,
\end{equation}
absent non-holomorphic Yukawa couplings, which relax the weight conditions by two units.
The length selection rule implies that this mixing first arises at two loops, where its only surviving contribution is built from four-dimensional tree amplitudes and tree form factors integrated over four-dimensional phase space.
Finite renormalizations that preserve the length selection rule cannot feed this entry through lower-order counterterms, so the zeros in \Eq{eq:dl1} remain unchanged under such changes of scheme.

The same logic extends to arbitrary loop order. At $L+1\ge 2$ loops with $\ell_i=\ell_j-L$, all contributions to $\gamma^{(L+1)}_{ij}$ proportional to an IR anomalous dimension, a $\beta$-function, or a lower-loop anomalous dimension, as well as every cut with a loop-level amplitude or form factor, are excluded by the leg counting used above, each requiring $\ell_i\ge\ell_j-L+1$ \cite{Bern:2019wie}. As a consequence, only the maximal $(L+2)$-particle cut of tree amplitudes and tree form factors survives, so that from Eq.~\eqref{eq:sewing} we obtain
\begin{equation}
\label{eq:gen}
    \gamma^{(L+1)}_{ij} = 0  \ \text{if}  \ \ell_i = \ell_j-L \ \land \  \big(w_i < w_j - 2L  \ \lor  \ \wb_i < \wb_j - 2L\big)
\end{equation}
with the last two conditions becoming $w_i < w_j - 2(L+1)$ and $\wb_i < \wb_j - 2(L+1)$ in the presence of non-holomorphic Yukawa couplings.
For $\ell_i<\ell_j-L$ even the maximal cut is unavailable, so $\gamma^{(L+1)}_{ij}=0$ \cite{Bern:2019wie}.
\smallskip

\pdfbookmark[1]{Scheme-dependent two-loop zeros}{Scheme-dependent two-loop zeros}
\paragraph{\textbf{Scheme-dependent two-loop zeros.}}

Subject to the complex-collinear factorization assumption discussed below, we find the following extension to $\ell_i \ge \ell_j$:
\begin{equation}
\label{eq:dl2}
 \hat\gamma^{(2)}_{ij} = 0  \ \ \text{if}  \ \ \ell_i \ge \ell_j \ \land \  \big(w_i < w_j - 4 \  \lor  \ \wb_i < \wb_j - 4 \big) \,.
\end{equation}
Here the hat denotes the \textit{holomorphic scheme}, in which these zeros hold even in the presence of non-holomorphic Yukawa couplings.
The key ingredient is the decomposition of the finite rational one-loop form factor, valid for $w(X)<w_j-2$ or $\wb(X)<\wb_j-2$,
\begin{equation}
\label{eq:contact}
F^{(1)}_j(X) = \textstyle\sum_i \kappa^{(1)}_{ij}\, F^{(0)}_i(X) \,,
\end{equation}
where $\kappa^{(1)}_{ij}$ are coefficients that depend only on the marginal couplings and the renormalization prescription, and the sum runs over dimension-$d$ operators with $\ell_j\le\ell_i\le d$ whose tree form factors are non-zero on $X$.
Removing these terms by the one-loop finite operator redefinition $\hat{\mathcal O}_j=\sum_i(\delta_{ij}-\kappa^{(1)}_{ij})\mathcal O_i$ defines the holomorphic scheme.

In a generic scheme, such as $\MSbar$, the entries in the region of \Eq{eq:dl2} are generally non-zero but satisfy
\begin{equation}
\label{eq:onedet}
    \gamma^{(2)}_{ij} = -[\kappa^{(1)},\gamma^{(1)}]_{ij} - \beta_g^{(1)}\partial_g \kappa^{(1)}_{ij} \,,
\end{equation}
with $[\kappa^{(1)},\gamma^{(1)}]_{ij}=\sum_k(\kappa^{(1)}_{ik}\gamma^{(1)}_{kj}-\gamma^{(1)}_{ik}\kappa^{(1)}_{kj})$.
Thus, this block carries no independent two-loop information: it is fixed entirely by the one-loop anomalous dimensions, $\beta$-functions, and coefficients $\kappa^{(1)}_{ij}$.
The latter are read off from \Eq{eq:contact} on the minimal target configurations $X_i$, so no two-loop integral is required.

The derivation of the existence of the holomorphic scheme is given in the Appendix. It assumes that the singularities of one-loop rational form factors in complex-collinear limits retain a factorized form compatible with the helicity-weight counting.
No general theorem establishes this factorization for complex momenta, where double and unreal poles can occur \cite{Bern:2005hs,Bern:2005ji,Bern:2005cq}, and Eqs.~\eqref{eq:dl2}--\eqref{eq:onedet} are therefore conditional on this assumption.
The scheme-independent two-loop zeros of \Eq{eq:dl1} and their higher-loop generalization in \Eq{eq:gen} rely only on tree-level data and do not depend on this assumption.

As a concrete application of these results, we summarize in Tables \ref{tab:dim5}, \ref{tab:dim6}, \ref{tab:dim7}, and \ref{tab:dim8} the structure that the selection rules of
Eqs.~\eqref{eq:gen} and \eqref{eq:dl2} impose on the anomalous-dimension matrix at operator
dimensions five, six, seven, and eight, respectively. \smallskip

\begin{table}[tbp]
    \centering
    \renewcommand{\arraystretch}{1.3}
    \begin{NiceTabular}{cc|ccc|ccc|c}
         & & $\phi F^2$ & $\psi^2 F$ & $\psi^2 \phi^2$ & $\phi \bar F^2$ & $\bar\psi^2 \bar F$ & $\bar\psi^2 \phi^2$ & $\phi^5$ \\
         & $(w,\wb)$ & $(1,5)$ & $(1,5)$ & $(3,5)$ & $(5,1)$ & $(5,1)$ & $(5,3)$ & $(5,5)$ \\ \hline
        $\phi F^2$ & $(1,5)$ & $(1)$ & $(1)$ & $(2)$ & $0_{(1)}$ & $0_{(1)}$ & \new $0_{(2)}$ & $(3)$ \\
        $\psi^2 F$ & $(1,5)$ & $(1)$ & $(1)$ & $(2)$ & $0_{(1)}$ & $0_{(1)}$ & \new $0_{(2)}$ & \new $\times_{(3)}$ \\
        $\psi^2 \phi^2$ & $(3,5)$ & $(1)$ & $(1)$ & $(1)$ & $0_{(1)}$ & $0_{(1)}$ & $y^2_{(1)}$ & $(2)$ \\ \hline
        $\phi \bar F^2$ & $(5,1)$ & $0_{(1)}$ & $0_{(1)}$ & \new $0_{(2)}$ & $(1)$ & $(1)$ & $(2)$ & $(3)$ \\
        $\bar\psi^2 \bar F$ & $(5,1)$ & $0_{(1)}$ & $0_{(1)}$ & \new $0_{(2)}$ & $(1)$ & $(1)$ & $(2)$ & \new $\times_{(3)}$ \\
        $\bar\psi^2 \phi^2$ & $(5,3)$ & $0_{(1)}$ & $0_{(1)}$ & $\bar y^2_{(1)}$ & $(1)$ & $(1)$ & $(1)$ & $(2)$ \\ \hline
        $\phi^5$ & $(5,5)$ & $(1)$ & $\times_{(1)}$ & $(1)$ & $(1)$ & $\times_{(1)}$ & $(1)$ & $(1)$ \\
    \end{NiceTabular}
    \caption{Structure of the anomalous-dimension matrix for dimension-five operators. The entry in row $\mathcal O_i$ and column $\mathcal O_j$ refers to $\gamma^{(L)}_{ij}$ at the loop order $L=\max(1,1+\ell_j-\ell_i)$, the lowest order allowed by the length selection rule \cite{Bern:2019wie}. At that order the mixing is determined by $(L+1)$-particle cuts of tree-level objects. Entries $0_{(L)}$ vanish by the helicity selection rule of \Eq{eq:gen}. Entries $y^2_{(L)}$ and $\bar y^2_{(L)}$ vanish as well unless non-holomorphic Yukawa couplings are present. $\times_{(L)}$ marks entries that are not forbidden by \Eq{eq:gen} but for which no $(L+1)$-particle cut exists. $(L)$ denotes entries that need not vanish. Moreover, cells shaded in gray vanish at two loops in the holomorphic scheme, as implied by \Eq{eq:dl2} (first appearing at dimension six), while those shaded in blue correspond to previously unknown scheme-independent zeros.}
    \label{tab:dim5}
\end{table}
\begin{table*}[tbp]
    \centering
    \renewcommand{\arraystretch}{1.3}
    \begin{NiceTabular}{cc|ccccc|ccccc|cccc}
         & & $F^3$ & $\phi^2 F^2$ & $\psi^2 \phi F$ & $\psi^4$ & $\psi^2 \phi^3$ & $\bar F^3$ & $\phi^2 \bar F^2$ & $\bar\psi^2 \phi \bar F$ & $\bar\psi^4$ & $\bar\psi^2 \phi^3$ & $\bar\psi^2 \psi^2$ & $D \bar\psi \psi \phi^2$ & $D^2 \phi^4$ & $\phi^6$ \\
         & $(w,\wb)$ & $(0,6)$ & $(2,6)$ & $(2,6)$ & $(2,6)$ & $(4,6)$ & $(6,0)$ & $(6,2)$ & $(6,2)$ & $(6,2)$ & $(6,4)$ & $(4,4)$ & $(4,4)$ & $(4,4)$ & $(6,6)$ \\ \hline
        $F^3$ & $(0,6)$ & $(1)$ & $(2)$ & $(2)$ & \new $\times_{(2)}$ & \new $\times_{(3)}$ & \ccol$0_{(1)}$ & \new $0_{(2)}$ & \new $0_{(2)}$ & \new $0_{(2)}$ & \new $0_{(3)}$ & \new $0_{(2)}$ & \new $0_{(2)}$ & \new $0_{(2)}$ & \new $\times_{(4)}$ \\
        $\phi^2 F^2$ & $(2,6)$ & $(1)$ & $(1)$ & $(1)$ & $\times_{(1)}$ & $(2)$ & $0_{(1)}$ & $0_{(1)}$ & $0_{(1)}$ & $0_{(1)}$ & \new $0_{(2)}$ & $0_{(1)}$ & $0_{(1)}$ & $0_{(1)}$ & $(3)$ \\
        $\psi^2 \phi F$ & $(2,6)$ & $(1)$ & $(1)$ & $(1)$ & $(1)$ & $(2)$ & $0_{(1)}$ & $0_{(1)}$ & $0_{(1)}$ & $0_{(1)}$ & \new $0_{(2)}$ & $0_{(1)}$ & $0_{(1)}$ & $0_{(1)}$ & \new $\times_{(3)}$ \\
        $\psi^4$ & $(2,6)$ & $\times_{(1)}$ & $\times_{(1)}$ & $(1)$ & $(1)$ & $(2)$ & $0_{(1)}$ & $0_{(1)}$ & $0_{(1)}$ & $0_{(1)}$ & \new $0_{(2)}$ & $y^2_{(1)}$ & $0_{(1)}$ & $0_{(1)}$ & \new $\times_{(3)}$ \\
        $\psi^2 \phi^3$ & $(4,6)$ & $\times_{(1)}$ & $(1)$ & $(1)$ & $(1)$ & $(1)$ & $0_{(1)}$ & $0_{(1)}$ & $0_{(1)}$ & $0_{(1)}$ & $y^2_{(1)}$ & $(1)$ & $(1)$ & $(1)$ & $(2)$ \\ \hline
        $\bar F^3$ & $(6,0)$ & \ccol$0_{(1)}$ & \new $0_{(2)}$ & \new $0_{(2)}$ & \new $0_{(2)}$ & \new $0_{(3)}$ & $(1)$ & $(2)$ & $(2)$ & \new $\times_{(2)}$ & \new $\times_{(3)}$ & \new $0_{(2)}$ & \new $0_{(2)}$ & \new $0_{(2)}$ & \new $\times_{(4)}$ \\
        $\phi^2 \bar F^2$ & $(6,2)$ & $0_{(1)}$ & $0_{(1)}$ & $0_{(1)}$ & $0_{(1)}$ & \new $0_{(2)}$ & $(1)$ & $(1)$ & $(1)$ & $\times_{(1)}$ & $(2)$ & $0_{(1)}$ & $0_{(1)}$ & $0_{(1)}$ & $(3)$ \\
        $\bar\psi^2 \phi \bar F$ & $(6,2)$ & $0_{(1)}$ & $0_{(1)}$ & $0_{(1)}$ & $0_{(1)}$ & \new $0_{(2)}$ & $(1)$ & $(1)$ & $(1)$ & $(1)$ & $(2)$ & $0_{(1)}$ & $0_{(1)}$ & $0_{(1)}$ & \new $\times_{(3)}$ \\
        $\bar\psi^4$ & $(6,2)$ & $0_{(1)}$ & $0_{(1)}$ & $0_{(1)}$ & $0_{(1)}$ & \new $0_{(2)}$ & $\times_{(1)}$ & $\times_{(1)}$ & $(1)$ & $(1)$ & $(2)$ & $\bar y^2_{(1)}$ & $0_{(1)}$ & $0_{(1)}$ & \new $\times_{(3)}$ \\
        $\bar\psi^2 \phi^3$ & $(6,4)$ & $0_{(1)}$ & $0_{(1)}$ & $0_{(1)}$ & $0_{(1)}$ & $\bar y^2_{(1)}$ & $\times_{(1)}$ & $(1)$ & $(1)$ & $(1)$ & $(1)$ & $(1)$ & $(1)$ & $(1)$ & $(2)$ \\ \hline
        $\bar\psi^2 \psi^2$ & $(4,4)$ & $0_{(1)}$ & $0_{(1)}$ & $0_{(1)}$ & $\bar y^2_{(1)}$ & $(2)$ & $0_{(1)}$ & $0_{(1)}$ & $0_{(1)}$ & $y^2_{(1)}$ & $(2)$ & $(1)$ & $(1)$ & $\times_{(1)}$ & \new $\times_{(3)}$ \\
        $D \bar\psi \psi \phi^2$ & $(4,4)$ & $0_{(1)}$ & $0_{(1)}$ & $0_{(1)}$ & $0_{(1)}$ & $(2)$ & $0_{(1)}$ & $0_{(1)}$ & $0_{(1)}$ & $0_{(1)}$ & $(2)$ & $(1)$ & $(1)$ & $(1)$ & $(3)$ \\
        $D^2 \phi^4$ & $(4,4)$ & $0_{(1)}$ & $0_{(1)}$ & $0_{(1)}$ & $0_{(1)}$ & $(2)$ & $0_{(1)}$ & $0_{(1)}$ & $0_{(1)}$ & $0_{(1)}$ & $(2)$ & $\times_{(1)}$ & $(1)$ & $(1)$ & $(3)$ \\
        $\phi^6$ & $(6,6)$ & $\times_{(1)}$ & $(1)$ & $\times_{(1)}$ & $\times_{(1)}$ & $(1)$ & $\times_{(1)}$ & $(1)$ & $\times_{(1)}$ & $\times_{(1)}$ & $(1)$ & $\times_{(1)}$ & $(1)$ & $(1)$ & $(1)$ \\
    \end{NiceTabular}
    \caption{Structure of the anomalous-dimension matrix for dimension-six operators. The notation is described in Table \ref{tab:dim5}.}
    \label{tab:dim6}
\end{table*}
\begin{table*}[tbp]
    \centering
    \renewcommand{\arraystretch}{1.3}
    \resizebox{\textwidth}{!}{
    \begin{NiceTabular}{cc|ccccccccccc|ccccccccccc|cccc}
         & & $\phi F^3$ & $\psi^2 F^2$ & $D^2 \phi^3 F$ & $D \bar\psi \psi \phi F$ & $\bar\psi^2 F^2$ & $D^2 \psi^2 \phi^2$ & $D \bar\psi \psi^3$ & $\phi^3 F^2$ & $\psi^4 \phi$ & $\psi^2 \phi^2 F$ & $\psi^2 \phi^4$ & $\phi \bar F^3$ & $\bar\psi^2 \bar F^2$ & $D^2 \phi^3 \bar F$ & $D \bar\psi \psi \phi \bar F$ & $\psi^2 \bar F^2$ & $D^2 \bar\psi^2 \phi^2$ & $D \bar\psi^3 \psi$ & $\phi^3 \bar F^2$ & $\bar\psi^4 \phi$ & $\bar\psi^2 \phi^2 \bar F$ & $\bar\psi^2 \phi^4$ & $D^2 \phi^5$ & $D \bar\psi \psi \phi^3$ & $\bar\psi^2 \psi^2 \phi$ & $\phi^7$ \\
         & $(w,\wb)$ & $(1,7)$ & $(1,7)$ & $(3,5)$ & $(3,5)$ & $(3,5)$ & $(3,5)$ & $(3,5)$ & $(3,7)$ & $(3,7)$ & $(3,7)$ & $(5,7)$ & $(7,1)$ & $(7,1)$ & $(5,3)$ & $(5,3)$ & $(5,3)$ & $(5,3)$ & $(5,3)$ & $(7,3)$ & $(7,3)$ & $(7,3)$ & $(7,5)$ & $(5,5)$ & $(5,5)$ & $(5,5)$ & $(7,7)$ \\ \hline
        $\phi F^3$ & $(1,7)$ & $(1)$ & $(1)$ & $0_{(1)}$ & $0_{(1)}$ & $0_{(1)}$ & $0_{(1)}$ & $0_{(1)}$ & $(2)$ & \new $\times_{(2)}$ & $(2)$ & \new $\times_{(3)}$ & \ccol$0_{(1)}$ & \ccol$0_{(1)}$ & $0_{(1)}$ & $0_{(1)}$ & $0_{(1)}$ & $0_{(1)}$ & $0_{(1)}$ & \new $0_{(2)}$ & \new $0_{(2)}$ & \new $0_{(2)}$ & \new $0_{(3)}$ & \new $0_{(2)}$ & \new $0_{(2)}$ & \new $0_{(2)}$ & \new $\times_{(4)}$ \\
        $\psi^2 F^2$ & $(1,7)$ & $(1)$ & $(1)$ & $0_{(1)}$ & $0_{(1)}$ & $y^2_{(1)}$ & $0_{(1)}$ & $0_{(1)}$ & $(2)$ & $(2)$ & $(2)$ & $(3)$ & \ccol$0_{(1)}$ & \ccol$0_{(1)}$ & $0_{(1)}$ & $0_{(1)}$ & $0_{(1)}$ & $0_{(1)}$ & $0_{(1)}$ & \new $0_{(2)}$ & \new $0_{(2)}$ & \new $0_{(2)}$ & \new $0_{(3)}$ & \new $0_{(2)}$ & \new $0_{(2)}$ & \new $0_{(2)}$ & \new $\times_{(4)}$ \\
        $D^2 \phi^3 F$ & $(3,5)$ & $0_{(1)}$ & $0_{(1)}$ & $(1)$ & $(1)$ & $\times_{(1)}$ & $(1)$ & $\times_{(1)}$ & $(2)$ & \new $\times_{(2)}$ & $(2)$ & $(3)$ & $0_{(1)}$ & $0_{(1)}$ & $0_{(1)}$ & $0_{(1)}$ & $0_{(1)}$ & $0_{(1)}$ & $0_{(1)}$ & \new $0_{(2)}$ & \new $0_{(2)}$ & \new $0_{(2)}$ & $(3)$ & $(2)$ & $(2)$ & \new $\times_{(2)}$ & $(4)$ \\
        $D \bar\psi \psi \phi F$ & $(3,5)$ & $0_{(1)}$ & $0_{(1)}$ & $(1)$ & $(1)$ & $(1)$ & $(1)$ & $(1)$ & $(2)$ & $(2)$ & $(2)$ & $(3)$ & $0_{(1)}$ & $0_{(1)}$ & $0_{(1)}$ & $0_{(1)}$ & $0_{(1)}$ & $0_{(1)}$ & $0_{(1)}$ & \new $0_{(2)}$ & \new $y^2_{(2)}$ & \new $0_{(2)}$ & $(3)$ & \new $\times_{(2)}$ & $(2)$ & $(2)$ & \new $\times_{(4)}$ \\
        $\bar\psi^2 F^2$ & $(3,5)$ & $0_{(1)}$ & $\bar y^2_{(1)}$ & $\times_{(1)}$ & $(1)$ & $(1)$ & $\times_{(1)}$ & $\times_{(1)}$ & $(2)$ & \new $\times_{(2)}$ & \new $\times_{(2)}$ & \new $\times_{(3)}$ & $0_{(1)}$ & $0_{(1)}$ & $0_{(1)}$ & $0_{(1)}$ & $0_{(1)}$ & $0_{(1)}$ & $0_{(1)}$ & \new $0_{(2)}$ & \new $0_{(2)}$ & \new $0_{(2)}$ & $(3)$ & \new $\times_{(2)}$ & \new $\times_{(2)}$ & $(2)$ & \new $\times_{(4)}$ \\
        $D^2 \psi^2 \phi^2$ & $(3,5)$ & $0_{(1)}$ & $0_{(1)}$ & $(1)$ & $(1)$ & $\times_{(1)}$ & $(1)$ & $(1)$ & $(2)$ & $(2)$ & $(2)$ & $(3)$ & $0_{(1)}$ & $0_{(1)}$ & $0_{(1)}$ & $0_{(1)}$ & $0_{(1)}$ & $y^2_{(1)}$ & $0_{(1)}$ & \new $0_{(2)}$ & \new $0_{(2)}$ & \new $y^2_{(2)}$ & $(3)$ & $(2)$ & $(2)$ & $(2)$ & $(4)$ \\
        $D \bar\psi \psi^3$ & $(3,5)$ & $0_{(1)}$ & $0_{(1)}$ & $\times_{(1)}$ & $(1)$ & $\times_{(1)}$ & $(1)$ & $(1)$ & \new $\times_{(2)}$ & $(2)$ & $(2)$ & $(3)$ & $0_{(1)}$ & $0_{(1)}$ & $0_{(1)}$ & $0_{(1)}$ & $0_{(1)}$ & $0_{(1)}$ & $y^2_{(1)}$ & \new $0_{(2)}$ & \new $0_{(2)}$ & \new $0_{(2)}$ & \new $\times_{(3)}$ & \new $\times_{(2)}$ & $(2)$ & $(2)$ & \new $\times_{(4)}$ \\
        $\phi^3 F^2$ & $(3,7)$ & $(1)$ & $(1)$ & $(1)$ & $(1)$ & $(1)$ & $(1)$ & $\times_{(1)}$ & $(1)$ & $\times_{(1)}$ & $(1)$ & $(2)$ & $0_{(1)}$ & $0_{(1)}$ & $0_{(1)}$ & $0_{(1)}$ & $0_{(1)}$ & $0_{(1)}$ & $0_{(1)}$ & $0_{(1)}$ & $0_{(1)}$ & $0_{(1)}$ & \new $0_{(2)}$ & $0_{(1)}$ & $0_{(1)}$ & $0_{(1)}$ & $(3)$ \\
        $\psi^4 \phi$ & $(3,7)$ & $\times_{(1)}$ & $(1)$ & $\times_{(1)}$ & $(1)$ & $\times_{(1)}$ & $(1)$ & $(1)$ & $\times_{(1)}$ & $(1)$ & $(1)$ & $(2)$ & $0_{(1)}$ & $0_{(1)}$ & $0_{(1)}$ & $y^2_{(1)}$ & $0_{(1)}$ & $0_{(1)}$ & $0_{(1)}$ & $0_{(1)}$ & $0_{(1)}$ & $0_{(1)}$ & \new $0_{(2)}$ & $0_{(1)}$ & $0_{(1)}$ & $y^2_{(1)}$ & \new $\times_{(3)}$ \\
        $\psi^2 \phi^2 F$ & $(3,7)$ & $(1)$ & $(1)$ & $(1)$ & $(1)$ & $\times_{(1)}$ & $(1)$ & $(1)$ & $(1)$ & $(1)$ & $(1)$ & $(2)$ & $0_{(1)}$ & $0_{(1)}$ & $0_{(1)}$ & $0_{(1)}$ & $0_{(1)}$ & $y^2_{(1)}$ & $0_{(1)}$ & $0_{(1)}$ & $0_{(1)}$ & $0_{(1)}$ & \new $0_{(2)}$ & $0_{(1)}$ & $0_{(1)}$ & $0_{(1)}$ & \new $\times_{(3)}$ \\
        $\psi^2 \phi^4$ & $(5,7)$ & $\times_{(1)}$ & $(1)$ & $(1)$ & $(1)$ & $\times_{(1)}$ & $(1)$ & $(1)$ & $(1)$ & $(1)$ & $(1)$ & $(1)$ & $0_{(1)}$ & $0_{(1)}$ & $(1)$ & $(1)$ & $(1)$ & $(1)$ & $\times_{(1)}$ & $0_{(1)}$ & $0_{(1)}$ & $0_{(1)}$ & $y^2_{(1)}$ & $(1)$ & $(1)$ & $(1)$ & $(2)$ \\ \hline
        $\phi \bar F^3$ & $(7,1)$ & \ccol$0_{(1)}$ & \ccol$0_{(1)}$ & $0_{(1)}$ & $0_{(1)}$ & $0_{(1)}$ & $0_{(1)}$ & $0_{(1)}$ & \new $0_{(2)}$ & \new $0_{(2)}$ & \new $0_{(2)}$ & \new $0_{(3)}$ & $(1)$ & $(1)$ & $0_{(1)}$ & $0_{(1)}$ & $0_{(1)}$ & $0_{(1)}$ & $0_{(1)}$ & $(2)$ & \new $\times_{(2)}$ & $(2)$ & \new $\times_{(3)}$ & \new $0_{(2)}$ & \new $0_{(2)}$ & \new $0_{(2)}$ & \new $\times_{(4)}$ \\
        $\bar\psi^2 \bar F^2$ & $(7,1)$ & \ccol$0_{(1)}$ & \ccol$0_{(1)}$ & $0_{(1)}$ & $0_{(1)}$ & $0_{(1)}$ & $0_{(1)}$ & $0_{(1)}$ & \new $0_{(2)}$ & \new $0_{(2)}$ & \new $0_{(2)}$ & \new $0_{(3)}$ & $(1)$ & $(1)$ & $0_{(1)}$ & $0_{(1)}$ & $\bar y^2_{(1)}$ & $0_{(1)}$ & $0_{(1)}$ & $(2)$ & $(2)$ & $(2)$ & $(3)$ & \new $0_{(2)}$ & \new $0_{(2)}$ & \new $0_{(2)}$ & \new $\times_{(4)}$ \\
        $D^2 \phi^3 \bar F$ & $(5,3)$ & $0_{(1)}$ & $0_{(1)}$ & $0_{(1)}$ & $0_{(1)}$ & $0_{(1)}$ & $0_{(1)}$ & $0_{(1)}$ & \new $0_{(2)}$ & \new $0_{(2)}$ & \new $0_{(2)}$ & $(3)$ & $0_{(1)}$ & $0_{(1)}$ & $(1)$ & $(1)$ & $\times_{(1)}$ & $(1)$ & $\times_{(1)}$ & $(2)$ & \new $\times_{(2)}$ & $(2)$ & $(3)$ & $(2)$ & $(2)$ & \new $\times_{(2)}$ & $(4)$ \\
        $D \bar\psi \psi \phi \bar F$ & $(5,3)$ & $0_{(1)}$ & $0_{(1)}$ & $0_{(1)}$ & $0_{(1)}$ & $0_{(1)}$ & $0_{(1)}$ & $0_{(1)}$ & \new $0_{(2)}$ & \new $\bar y^2_{(2)}$ & \new $0_{(2)}$ & $(3)$ & $0_{(1)}$ & $0_{(1)}$ & $(1)$ & $(1)$ & $(1)$ & $(1)$ & $(1)$ & $(2)$ & $(2)$ & $(2)$ & $(3)$ & \new $\times_{(2)}$ & $(2)$ & $(2)$ & \new $\times_{(4)}$ \\
        $\psi^2 \bar F^2$ & $(5,3)$ & $0_{(1)}$ & $0_{(1)}$ & $0_{(1)}$ & $0_{(1)}$ & $0_{(1)}$ & $0_{(1)}$ & $0_{(1)}$ & \new $0_{(2)}$ & \new $0_{(2)}$ & \new $0_{(2)}$ & $(3)$ & $0_{(1)}$ & $y^2_{(1)}$ & $\times_{(1)}$ & $(1)$ & $(1)$ & $\times_{(1)}$ & $\times_{(1)}$ & $(2)$ & \new $\times_{(2)}$ & \new $\times_{(2)}$ & \new $\times_{(3)}$ & \new $\times_{(2)}$ & \new $\times_{(2)}$ & $(2)$ & \new $\times_{(4)}$ \\
        $D^2 \bar\psi^2 \phi^2$ & $(5,3)$ & $0_{(1)}$ & $0_{(1)}$ & $0_{(1)}$ & $0_{(1)}$ & $0_{(1)}$ & $\bar y^2_{(1)}$ & $0_{(1)}$ & \new $0_{(2)}$ & \new $0_{(2)}$ & \new $\bar y^2_{(2)}$ & $(3)$ & $0_{(1)}$ & $0_{(1)}$ & $(1)$ & $(1)$ & $\times_{(1)}$ & $(1)$ & $(1)$ & $(2)$ & $(2)$ & $(2)$ & $(3)$ & $(2)$ & $(2)$ & $(2)$ & $(4)$ \\
        $D \bar\psi^3 \psi$ & $(5,3)$ & $0_{(1)}$ & $0_{(1)}$ & $0_{(1)}$ & $0_{(1)}$ & $0_{(1)}$ & $0_{(1)}$ & $\bar y^2_{(1)}$ & \new $0_{(2)}$ & \new $0_{(2)}$ & \new $0_{(2)}$ & \new $\times_{(3)}$ & $0_{(1)}$ & $0_{(1)}$ & $\times_{(1)}$ & $(1)$ & $\times_{(1)}$ & $(1)$ & $(1)$ & \new $\times_{(2)}$ & $(2)$ & $(2)$ & $(3)$ & \new $\times_{(2)}$ & $(2)$ & $(2)$ & \new $\times_{(4)}$ \\
        $\phi^3 \bar F^2$ & $(7,3)$ & $0_{(1)}$ & $0_{(1)}$ & $0_{(1)}$ & $0_{(1)}$ & $0_{(1)}$ & $0_{(1)}$ & $0_{(1)}$ & $0_{(1)}$ & $0_{(1)}$ & $0_{(1)}$ & \new $0_{(2)}$ & $(1)$ & $(1)$ & $(1)$ & $(1)$ & $(1)$ & $(1)$ & $\times_{(1)}$ & $(1)$ & $\times_{(1)}$ & $(1)$ & $(2)$ & $0_{(1)}$ & $0_{(1)}$ & $0_{(1)}$ & $(3)$ \\
        $\bar\psi^4 \phi$ & $(7,3)$ & $0_{(1)}$ & $0_{(1)}$ & $0_{(1)}$ & $\bar y^2_{(1)}$ & $0_{(1)}$ & $0_{(1)}$ & $0_{(1)}$ & $0_{(1)}$ & $0_{(1)}$ & $0_{(1)}$ & \new $0_{(2)}$ & $\times_{(1)}$ & $(1)$ & $\times_{(1)}$ & $(1)$ & $\times_{(1)}$ & $(1)$ & $(1)$ & $\times_{(1)}$ & $(1)$ & $(1)$ & $(2)$ & $0_{(1)}$ & $0_{(1)}$ & $\bar y^2_{(1)}$ & \new $\times_{(3)}$ \\
        $\bar\psi^2 \phi^2 \bar F$ & $(7,3)$ & $0_{(1)}$ & $0_{(1)}$ & $0_{(1)}$ & $0_{(1)}$ & $0_{(1)}$ & $\bar y^2_{(1)}$ & $0_{(1)}$ & $0_{(1)}$ & $0_{(1)}$ & $0_{(1)}$ & \new $0_{(2)}$ & $(1)$ & $(1)$ & $(1)$ & $(1)$ & $\times_{(1)}$ & $(1)$ & $(1)$ & $(1)$ & $(1)$ & $(1)$ & $(2)$ & $0_{(1)}$ & $0_{(1)}$ & $0_{(1)}$ & \new $\times_{(3)}$ \\
        $\bar\psi^2 \phi^4$ & $(7,5)$ & $0_{(1)}$ & $0_{(1)}$ & $(1)$ & $(1)$ & $(1)$ & $(1)$ & $\times_{(1)}$ & $0_{(1)}$ & $0_{(1)}$ & $0_{(1)}$ & $\bar y^2_{(1)}$ & $\times_{(1)}$ & $(1)$ & $(1)$ & $(1)$ & $\times_{(1)}$ & $(1)$ & $(1)$ & $(1)$ & $(1)$ & $(1)$ & $(1)$ & $(1)$ & $(1)$ & $(1)$ & $(2)$ \\ \hline
        $D^2 \phi^5$ & $(5,5)$ & $0_{(1)}$ & $0_{(1)}$ & $(1)$ & $\times_{(1)}$ & $\times_{(1)}$ & $(1)$ & $\times_{(1)}$ & $0_{(1)}$ & $0_{(1)}$ & $0_{(1)}$ & $(2)$ & $0_{(1)}$ & $0_{(1)}$ & $(1)$ & $\times_{(1)}$ & $\times_{(1)}$ & $(1)$ & $\times_{(1)}$ & $0_{(1)}$ & $0_{(1)}$ & $0_{(1)}$ & $(2)$ & $(1)$ & $(1)$ & $\times_{(1)}$ & $(3)$ \\
        $D \bar\psi \psi \phi^3$ & $(5,5)$ & $0_{(1)}$ & $0_{(1)}$ & $(1)$ & $(1)$ & $\times_{(1)}$ & $(1)$ & $(1)$ & $0_{(1)}$ & $0_{(1)}$ & $0_{(1)}$ & $(2)$ & $0_{(1)}$ & $0_{(1)}$ & $(1)$ & $(1)$ & $\times_{(1)}$ & $(1)$ & $(1)$ & $0_{(1)}$ & $0_{(1)}$ & $0_{(1)}$ & $(2)$ & $(1)$ & $(1)$ & $(1)$ & $(3)$ \\
        $\bar\psi^2 \psi^2 \phi$ & $(5,5)$ & $0_{(1)}$ & $0_{(1)}$ & $\times_{(1)}$ & $(1)$ & $(1)$ & $(1)$ & $(1)$ & $0_{(1)}$ & $\bar y^2_{(1)}$ & $0_{(1)}$ & $(2)$ & $0_{(1)}$ & $0_{(1)}$ & $\times_{(1)}$ & $(1)$ & $(1)$ & $(1)$ & $(1)$ & $0_{(1)}$ & $y^2_{(1)}$ & $0_{(1)}$ & $(2)$ & $\times_{(1)}$ & $(1)$ & $(1)$ & \new $\times_{(3)}$ \\
        $\phi^7$ & $(7,7)$ & $\times_{(1)}$ & $\times_{(1)}$ & $(1)$ & $\times_{(1)}$ & $\times_{(1)}$ & $(1)$ & $\times_{(1)}$ & $(1)$ & $\times_{(1)}$ & $\times_{(1)}$ & $(1)$ & $\times_{(1)}$ & $\times_{(1)}$ & $(1)$ & $\times_{(1)}$ & $\times_{(1)}$ & $(1)$ & $\times_{(1)}$ & $(1)$ & $\times_{(1)}$ & $\times_{(1)}$ & $(1)$ & $(1)$ & $(1)$ & $\times_{(1)}$ & $(1)$ \\
    \end{NiceTabular}
    }
    \caption{Structure of the anomalous-dimension matrix for dimension-seven operators. The notation is described in Table \ref{tab:dim5}.}
    \label{tab:dim7}
\end{table*}
\begin{turnpage}
\begin{table*}[h!t!b!]
    \centering
    \renewcommand{\arraystretch}{1.3}
    \resizebox{\textheight}{!}{
    \begin{NiceTabular}{cc|cccccccccccccccccc|cccccccccccccccccc|cccccccccccc}
         & & $F^4$ & $D^2 \phi^2 F^2$ & $D^2 \psi^2 \phi F$ & $D^2 \psi^4$ & $D \bar\psi \psi F^2$ & $\phi^2 F^3$ & $\psi^2 \phi F^2$ & $\psi^4 F$ & $D^2 \phi^4 F$ & $D^2 \psi^2 \phi^3$ & $D \bar\psi \psi \phi^2 F$ & $D \bar\psi \psi^3 \phi$ & $\bar\psi^2 \phi F^2$ & $\bar\psi^2 \psi^2 F$ & $\phi^4 F^2$ & $\psi^2 \phi^3 F$ & $\psi^4 \phi^2$ & $\psi^2 \phi^5$ & $\bar F^4$ & $D^2 \phi^2 \bar F^2$ & $D^2 \bar\psi^2 \phi \bar F$ & $D^2 \bar\psi^4$ & $D \bar\psi \psi \bar F^2$ & $\phi^2 \bar F^3$ & $\bar\psi^2 \phi \bar F^2$ & $\bar\psi^4 \bar F$ & $D^2 \phi^4 \bar F$ & $D^2 \bar\psi^2 \phi^3$ & $D \bar\psi \psi \phi^2 \bar F$ & $D \bar\psi^3 \psi \phi$ & $\psi^2 \phi \bar F^2$ & $\bar\psi^2 \psi^2 \bar F$ & $\phi^4 \bar F^2$ & $\bar\psi^2 \phi^3 \bar F$ & $\bar\psi^4 \phi^2$ & $\bar\psi^2 \phi^5$ & $F^2 \bar F^2$ & $D^2 \phi^2 F \bar F$ & $D^4 \phi^4$ & $D^2 \psi^2 \phi \bar F$ & $D^2 \bar\psi^2 \phi F$ & $D \bar\psi \psi F \bar F$ & $D^3 \bar\psi \psi \phi^2$ & $D^2 \bar\psi^2 \psi^2$ & $D^2 \phi^6$ & $D \bar\psi \psi \phi^4$ & $\bar\psi^2 \psi^2 \phi^2$ & $\phi^8$ \\
         & $(w,\wb)$ & $(0,8)$ & $(2,6)$ & $(2,6)$ & $(2,6)$ & $(2,6)$ & $(2,8)$ & $(2,8)$ & $(2,8)$ & $(4,6)$ & $(4,6)$ & $(4,6)$ & $(4,6)$ & $(4,6)$ & $(4,6)$ & $(4,8)$ & $(4,8)$ & $(4,8)$ & $(6,8)$ & $(8,0)$ & $(6,2)$ & $(6,2)$ & $(6,2)$ & $(6,2)$ & $(8,2)$ & $(8,2)$ & $(8,2)$ & $(6,4)$ & $(6,4)$ & $(6,4)$ & $(6,4)$ & $(6,4)$ & $(6,4)$ & $(8,4)$ & $(8,4)$ & $(8,4)$ & $(8,6)$ & $(4,4)$ & $(4,4)$ & $(4,4)$ & $(4,4)$ & $(4,4)$ & $(4,4)$ & $(4,4)$ & $(4,4)$ & $(6,6)$ & $(6,6)$ & $(6,6)$ & $(8,8)$ \\ \hline
        $F^4$ & $(0,8)$ & $(1)$ & $0_{(1)}$ & $0_{(1)}$ & $0_{(1)}$ & $0_{(1)}$ & $(2)$ & $(2)$ & \new $\times_{(2)}$ & \new $0_{(2)}$ & \new $0_{(2)}$ & \new $0_{(2)}$ & \new $0_{(2)}$ & \new $0_{(2)}$ & \new $0_{(2)}$ & $(3)$ & \new $\times_{(3)}$ & \new $\times_{(3)}$ & \new $\times_{(4)}$ & \ccol$0_{(1)}$ & \ccol$0_{(1)}$ & \ccol$0_{(1)}$ & \ccol$0_{(1)}$ & \ccol$0_{(1)}$ & \new $0_{(2)}$ & \new $0_{(2)}$ & \new $0_{(2)}$ & \new $0_{(2)}$ & \new $0_{(2)}$ & \new $0_{(2)}$ & \new $0_{(2)}$ & \new $0_{(2)}$ & \new $0_{(2)}$ & \new $0_{(3)}$ & \new $0_{(3)}$ & \new $0_{(3)}$ & \new $0_{(4)}$ & $0_{(1)}$ & $0_{(1)}$ & $0_{(1)}$ & $0_{(1)}$ & $0_{(1)}$ & $0_{(1)}$ & $0_{(1)}$ & $0_{(1)}$ & \new $0_{(3)}$ & \new $0_{(3)}$ & \new $0_{(3)}$ & \new $\times_{(5)}$ \\
        $D^2 \phi^2 F^2$ & $(2,6)$ & $0_{(1)}$ & $(1)$ & $(1)$ & $\times_{(1)}$ & $(1)$ & $(2)$ & $(2)$ & \new $\times_{(2)}$ & $(2)$ & $(2)$ & $(2)$ & \new $\times_{(2)}$ & $(2)$ & \new $\times_{(2)}$ & $(3)$ & $(3)$ & $(3)$ & $(4)$ & \ccol$0_{(1)}$ & $0_{(1)}$ & $0_{(1)}$ & $0_{(1)}$ & $0_{(1)}$ & \new $0_{(2)}$ & \new $0_{(2)}$ & \new $0_{(2)}$ & \new $0_{(2)}$ & \new $0_{(2)}$ & \new $0_{(2)}$ & \new $0_{(2)}$ & \new $0_{(2)}$ & \new $0_{(2)}$ & \new $0_{(3)}$ & \new $0_{(3)}$ & \new $y^2_{(3)}$ & $(4)$ & $0_{(1)}$ & $0_{(1)}$ & $0_{(1)}$ & $0_{(1)}$ & $0_{(1)}$ & $0_{(1)}$ & $0_{(1)}$ & $0_{(1)}$ & $(3)$ & $(3)$ & $(3)$ & $(5)$ \\
        $D^2 \psi^2 \phi F$ & $(2,6)$ & $0_{(1)}$ & $(1)$ & $(1)$ & $(1)$ & $(1)$ & $(2)$ & $(2)$ & $(2)$ & $(2)$ & $(2)$ & $(2)$ & $(2)$ & $(2)$ & $(2)$ & $(3)$ & $(3)$ & $(3)$ & $(4)$ & \ccol$0_{(1)}$ & $0_{(1)}$ & $0_{(1)}$ & $0_{(1)}$ & $0_{(1)}$ & \new $0_{(2)}$ & \new $0_{(2)}$ & \new $0_{(2)}$ & \new $0_{(2)}$ & \new $0_{(2)}$ & \new $0_{(2)}$ & \new $y^2_{(2)}$ & \new $0_{(2)}$ & \new $0_{(2)}$ & \new $0_{(3)}$ & \new $0_{(3)}$ & \new $0_{(3)}$ & \new $\times_{(4)}$ & $0_{(1)}$ & $0_{(1)}$ & $0_{(1)}$ & $0_{(1)}$ & $y^2_{(1)}$ & $0_{(1)}$ & $0_{(1)}$ & $0_{(1)}$ & \new $\times_{(3)}$ & $(3)$ & $(3)$ & \new $\times_{(5)}$ \\
        $D^2 \psi^4$ & $(2,6)$ & $0_{(1)}$ & $\times_{(1)}$ & $(1)$ & $(1)$ & $\times_{(1)}$ & \new $\times_{(2)}$ & $(2)$ & $(2)$ & \new $\times_{(2)}$ & $(2)$ & \new $\times_{(2)}$ & $(2)$ & \new $\times_{(2)}$ & $(2)$ & \new $\times_{(3)}$ & $(3)$ & $(3)$ & $(4)$ & \ccol$0_{(1)}$ & $0_{(1)}$ & $0_{(1)}$ & $0_{(1)}$ & $0_{(1)}$ & \new $0_{(2)}$ & \new $0_{(2)}$ & \new $0_{(2)}$ & \new $0_{(2)}$ & \new $0_{(2)}$ & \new $0_{(2)}$ & \new $0_{(2)}$ & \new $0_{(2)}$ & \new $y^2_{(2)}$ & \new $0_{(3)}$ & \new $0_{(3)}$ & \new $0_{(3)}$ & \new $\times_{(4)}$ & $0_{(1)}$ & $0_{(1)}$ & $0_{(1)}$ & $0_{(1)}$ & $0_{(1)}$ & $0_{(1)}$ & $0_{(1)}$ & $y^2_{(1)}$ & \new $\times_{(3)}$ & \new $\times_{(3)}$ & $(3)$ & \new $\times_{(5)}$ \\
        $D \bar\psi \psi F^2$ & $(2,6)$ & $0_{(1)}$ & $(1)$ & $(1)$ & $\times_{(1)}$ & $(1)$ & $(2)$ & $(2)$ & $(2)$ & \new $\times_{(2)}$ & \new $\times_{(2)}$ & $(2)$ & $(2)$ & $(2)$ & $(2)$ & $(3)$ & $(3)$ & \new $\times_{(3)}$ & \new $\times_{(4)}$ & \ccol$0_{(1)}$ & $0_{(1)}$ & $0_{(1)}$ & $0_{(1)}$ & $0_{(1)}$ & \new $0_{(2)}$ & \new $0_{(2)}$ & \new $0_{(2)}$ & \new $0_{(2)}$ & \new $0_{(2)}$ & \new $0_{(2)}$ & \new $0_{(2)}$ & \new $0_{(2)}$ & \new $0_{(2)}$ & \new $0_{(3)}$ & \new $0_{(3)}$ & \new $0_{(3)}$ & \new $\times_{(4)}$ & $0_{(1)}$ & $0_{(1)}$ & $0_{(1)}$ & $0_{(1)}$ & $0_{(1)}$ & $0_{(1)}$ & $0_{(1)}$ & $0_{(1)}$ & \new $\times_{(3)}$ & $(3)$ & $(3)$ & \new $\times_{(5)}$ \\
        $\phi^2 F^3$ & $(2,8)$ & $(1)$ & $(1)$ & $(1)$ & $\times_{(1)}$ & $(1)$ & $(1)$ & $(1)$ & $\times_{(1)}$ & $0_{(1)}$ & $0_{(1)}$ & $0_{(1)}$ & $0_{(1)}$ & $0_{(1)}$ & $0_{(1)}$ & $(2)$ & $(2)$ & \new $\times_{(2)}$ & \new $\times_{(3)}$ & \ccol$0_{(1)}$ & $0_{(1)}$ & $0_{(1)}$ & $0_{(1)}$ & $0_{(1)}$ & \ccol$0_{(1)}$ & \ccol$0_{(1)}$ & \ccol$0_{(1)}$ & $0_{(1)}$ & $0_{(1)}$ & $0_{(1)}$ & $0_{(1)}$ & $0_{(1)}$ & $0_{(1)}$ & \new $0_{(2)}$ & \new $0_{(2)}$ & \new $0_{(2)}$ & \new $0_{(3)}$ & $0_{(1)}$ & $0_{(1)}$ & $0_{(1)}$ & $0_{(1)}$ & $0_{(1)}$ & $0_{(1)}$ & $0_{(1)}$ & $0_{(1)}$ & \new $0_{(2)}$ & \new $0_{(2)}$ & \new $0_{(2)}$ & \new $\times_{(4)}$ \\
        $\psi^2 \phi F^2$ & $(2,8)$ & $(1)$ & $(1)$ & $(1)$ & $(1)$ & $(1)$ & $(1)$ & $(1)$ & $(1)$ & $0_{(1)}$ & $0_{(1)}$ & $0_{(1)}$ & $0_{(1)}$ & $y^2_{(1)}$ & $0_{(1)}$ & $(2)$ & $(2)$ & $(2)$ & $(3)$ & \ccol$0_{(1)}$ & $0_{(1)}$ & $0_{(1)}$ & $0_{(1)}$ & $0_{(1)}$ & \ccol$0_{(1)}$ & \ccol$0_{(1)}$ & \ccol$0_{(1)}$ & $0_{(1)}$ & $0_{(1)}$ & $0_{(1)}$ & $0_{(1)}$ & $0_{(1)}$ & $0_{(1)}$ & \new $0_{(2)}$ & \new $0_{(2)}$ & \new $0_{(2)}$ & \new $0_{(3)}$ & $0_{(1)}$ & $0_{(1)}$ & $0_{(1)}$ & $0_{(1)}$ & $y^2_{(1)}$ & $0_{(1)}$ & $0_{(1)}$ & $0_{(1)}$ & \new $0_{(2)}$ & \new $0_{(2)}$ & \new $0_{(2)}$ & \new $\times_{(4)}$ \\
        $\psi^4 F$ & $(2,8)$ & $\times_{(1)}$ & $\times_{(1)}$ & $(1)$ & $(1)$ & $(1)$ & $\times_{(1)}$ & $(1)$ & $(1)$ & $0_{(1)}$ & $0_{(1)}$ & $0_{(1)}$ & $0_{(1)}$ & $0_{(1)}$ & $y^2_{(1)}$ & \new $\times_{(2)}$ & $(2)$ & $(2)$ & \new $\times_{(3)}$ & \ccol$0_{(1)}$ & $0_{(1)}$ & $0_{(1)}$ & $0_{(1)}$ & $0_{(1)}$ & \ccol$0_{(1)}$ & \ccol$0_{(1)}$ & \ccol$0_{(1)}$ & $0_{(1)}$ & $0_{(1)}$ & $0_{(1)}$ & $0_{(1)}$ & $0_{(1)}$ & $0_{(1)}$ & \new $0_{(2)}$ & \new $0_{(2)}$ & \new $0_{(2)}$ & \new $0_{(3)}$ & $0_{(1)}$ & $0_{(1)}$ & $0_{(1)}$ & $0_{(1)}$ & $0_{(1)}$ & $y^2_{(1)}$ & $0_{(1)}$ & $y^2_{(1)}$ & \new $0_{(2)}$ & \new $0_{(2)}$ & \new $0_{(2)}$ & \new $\times_{(4)}$ \\
        $D^2 \phi^4 F$ & $(4,6)$ & $0_{(1)}$ & $(1)$ & $(1)$ & $\times_{(1)}$ & $\times_{(1)}$ & $0_{(1)}$ & $0_{(1)}$ & $0_{(1)}$ & $(1)$ & $(1)$ & $(1)$ & $\times_{(1)}$ & $\times_{(1)}$ & $\times_{(1)}$ & $(2)$ & $(2)$ & \new $\times_{(2)}$ & $(3)$ & $0_{(1)}$ & $0_{(1)}$ & $0_{(1)}$ & $0_{(1)}$ & $0_{(1)}$ & $0_{(1)}$ & $0_{(1)}$ & $0_{(1)}$ & $0_{(1)}$ & $0_{(1)}$ & $0_{(1)}$ & $0_{(1)}$ & $0_{(1)}$ & $0_{(1)}$ & \new $0_{(2)}$ & \new $0_{(2)}$ & \new $0_{(2)}$ & $(3)$ & $\times_{(1)}$ & $(1)$ & $(1)$ & $\times_{(1)}$ & $(1)$ & $\times_{(1)}$ & $(1)$ & $\times_{(1)}$ & $(2)$ & $(2)$ & \new $\times_{(2)}$ & $(4)$ \\
        $D^2 \psi^2 \phi^3$ & $(4,6)$ & $0_{(1)}$ & $(1)$ & $(1)$ & $(1)$ & $\times_{(1)}$ & $0_{(1)}$ & $0_{(1)}$ & $0_{(1)}$ & $(1)$ & $(1)$ & $(1)$ & $(1)$ & $\times_{(1)}$ & $\times_{(1)}$ & $(2)$ & $(2)$ & $(2)$ & $(3)$ & $0_{(1)}$ & $0_{(1)}$ & $0_{(1)}$ & $0_{(1)}$ & $0_{(1)}$ & $0_{(1)}$ & $0_{(1)}$ & $0_{(1)}$ & $0_{(1)}$ & $y^2_{(1)}$ & $0_{(1)}$ & $0_{(1)}$ & $0_{(1)}$ & $0_{(1)}$ & \new $0_{(2)}$ & \new $y^2_{(2)}$ & \new $0_{(2)}$ & $(3)$ & $\times_{(1)}$ & $(1)$ & $(1)$ & $(1)$ & $\times_{(1)}$ & $\times_{(1)}$ & $(1)$ & $(1)$ & $(2)$ & $(2)$ & $(2)$ & $(4)$ \\
        $D \bar\psi \psi \phi^2 F$ & $(4,6)$ & $0_{(1)}$ & $(1)$ & $(1)$ & $\times_{(1)}$ & $(1)$ & $0_{(1)}$ & $0_{(1)}$ & $0_{(1)}$ & $(1)$ & $(1)$ & $(1)$ & $(1)$ & $(1)$ & $(1)$ & $(2)$ & $(2)$ & $(2)$ & $(3)$ & $0_{(1)}$ & $0_{(1)}$ & $0_{(1)}$ & $0_{(1)}$ & $0_{(1)}$ & $0_{(1)}$ & $0_{(1)}$ & $0_{(1)}$ & $0_{(1)}$ & $0_{(1)}$ & $0_{(1)}$ & $0_{(1)}$ & $0_{(1)}$ & $0_{(1)}$ & \new $0_{(2)}$ & \new $0_{(2)}$ & \new $y^2_{(2)}$ & $(3)$ & $\times_{(1)}$ & $(1)$ & $(1)$ & $(1)$ & $(1)$ & $(1)$ & $(1)$ & $(1)$ & \new $\times_{(2)}$ & $(2)$ & $(2)$ & \new $\times_{(4)}$ \\
        $D \bar\psi \psi^3 \phi$ & $(4,6)$ & $0_{(1)}$ & $\times_{(1)}$ & $(1)$ & $(1)$ & $(1)$ & $0_{(1)}$ & $0_{(1)}$ & $0_{(1)}$ & $\times_{(1)}$ & $(1)$ & $(1)$ & $(1)$ & $\times_{(1)}$ & $(1)$ & \new $\times_{(2)}$ & $(2)$ & $(2)$ & $(3)$ & $0_{(1)}$ & $0_{(1)}$ & $y^2_{(1)}$ & $0_{(1)}$ & $0_{(1)}$ & $0_{(1)}$ & $0_{(1)}$ & $0_{(1)}$ & $0_{(1)}$ & $0_{(1)}$ & $0_{(1)}$ & $y^2_{(1)}$ & $0_{(1)}$ & $0_{(1)}$ & \new $0_{(2)}$ & \new $0_{(2)}$ & \new $0_{(2)}$ & \new $\times_{(3)}$ & $\times_{(1)}$ & $\times_{(1)}$ & $\times_{(1)}$ & $(1)$ & $(1)$ & $(1)$ & $(1)$ & $(1)$ & \new $\times_{(2)}$ & $(2)$ & $(2)$ & \new $\times_{(4)}$ \\
        $\bar\psi^2 \phi F^2$ & $(4,6)$ & $0_{(1)}$ & $(1)$ & $(1)$ & $\times_{(1)}$ & $(1)$ & $0_{(1)}$ & $\bar y^2_{(1)}$ & $0_{(1)}$ & $\times_{(1)}$ & $\times_{(1)}$ & $(1)$ & $\times_{(1)}$ & $(1)$ & $(1)$ & $(2)$ & \new $\times_{(2)}$ & \new $\times_{(2)}$ & \new $\times_{(3)}$ & $0_{(1)}$ & $0_{(1)}$ & $0_{(1)}$ & $0_{(1)}$ & $0_{(1)}$ & $0_{(1)}$ & $0_{(1)}$ & $0_{(1)}$ & $0_{(1)}$ & $0_{(1)}$ & $0_{(1)}$ & $0_{(1)}$ & $0_{(1)}$ & $0_{(1)}$ & \new $0_{(2)}$ & \new $0_{(2)}$ & \new $0_{(2)}$ & $(3)$ & $(1)$ & $(1)$ & $\times_{(1)}$ & $\times_{(1)}$ & $(1)$ & $(1)$ & $(1)$ & $(1)$ & \new $\times_{(2)}$ & \new $\times_{(2)}$ & $(2)$ & \new $\times_{(4)}$ \\
        $\bar\psi^2 \psi^2 F$ & $(4,6)$ & $0_{(1)}$ & $\times_{(1)}$ & $(1)$ & $(1)$ & $(1)$ & $0_{(1)}$ & $0_{(1)}$ & $\bar y^2_{(1)}$ & $\times_{(1)}$ & $\times_{(1)}$ & $(1)$ & $(1)$ & $(1)$ & $(1)$ & \new $\times_{(2)}$ & $(2)$ & \new $\times_{(2)}$ & \new $\times_{(3)}$ & $0_{(1)}$ & $0_{(1)}$ & $0_{(1)}$ & $y^2_{(1)}$ & $0_{(1)}$ & $0_{(1)}$ & $0_{(1)}$ & $0_{(1)}$ & $0_{(1)}$ & $0_{(1)}$ & $0_{(1)}$ & $0_{(1)}$ & $0_{(1)}$ & $0_{(1)}$ & \new $0_{(2)}$ & \new $0_{(2)}$ & \new $0_{(2)}$ & \new $\times_{(3)}$ & $\times_{(1)}$ & $\times_{(1)}$ & $\times_{(1)}$ & $(1)$ & $(1)$ & $(1)$ & $(1)$ & $(1)$ & \new $\times_{(2)}$ & \new $\times_{(2)}$ & $(2)$ & \new $\times_{(4)}$ \\
        $\phi^4 F^2$ & $(4,8)$ & $(1)$ & $(1)$ & $(1)$ & $\times_{(1)}$ & $(1)$ & $(1)$ & $(1)$ & $\times_{(1)}$ & $(1)$ & $(1)$ & $(1)$ & $\times_{(1)}$ & $(1)$ & $\times_{(1)}$ & $(1)$ & $(1)$ & $\times_{(1)}$ & $(2)$ & $0_{(1)}$ & $0_{(1)}$ & $0_{(1)}$ & $0_{(1)}$ & $0_{(1)}$ & $0_{(1)}$ & $0_{(1)}$ & $0_{(1)}$ & $0_{(1)}$ & $0_{(1)}$ & $0_{(1)}$ & $0_{(1)}$ & $0_{(1)}$ & $0_{(1)}$ & $0_{(1)}$ & $0_{(1)}$ & $0_{(1)}$ & \new $0_{(2)}$ & $(1)$ & $(1)$ & $(1)$ & $\times_{(1)}$ & $(1)$ & $\times_{(1)}$ & $(1)$ & $\times_{(1)}$ & $0_{(1)}$ & $0_{(1)}$ & $0_{(1)}$ & $(3)$ \\
        $\psi^2 \phi^3 F$ & $(4,8)$ & $\times_{(1)}$ & $(1)$ & $(1)$ & $(1)$ & $(1)$ & $(1)$ & $(1)$ & $(1)$ & $(1)$ & $(1)$ & $(1)$ & $(1)$ & $\times_{(1)}$ & $(1)$ & $(1)$ & $(1)$ & $(1)$ & $(2)$ & $0_{(1)}$ & $0_{(1)}$ & $0_{(1)}$ & $0_{(1)}$ & $0_{(1)}$ & $0_{(1)}$ & $0_{(1)}$ & $0_{(1)}$ & $0_{(1)}$ & $y^2_{(1)}$ & $0_{(1)}$ & $0_{(1)}$ & $0_{(1)}$ & $0_{(1)}$ & $0_{(1)}$ & $0_{(1)}$ & $0_{(1)}$ & \new $0_{(2)}$ & $\times_{(1)}$ & $(1)$ & $(1)$ & $(1)$ & $(1)$ & $(1)$ & $(1)$ & $(1)$ & $0_{(1)}$ & $0_{(1)}$ & $0_{(1)}$ & \new $\times_{(3)}$ \\
        $\psi^4 \phi^2$ & $(4,8)$ & $\times_{(1)}$ & $(1)$ & $(1)$ & $(1)$ & $\times_{(1)}$ & $\times_{(1)}$ & $(1)$ & $(1)$ & $\times_{(1)}$ & $(1)$ & $(1)$ & $(1)$ & $\times_{(1)}$ & $\times_{(1)}$ & $\times_{(1)}$ & $(1)$ & $(1)$ & $(2)$ & $0_{(1)}$ & $y^2_{(1)}$ & $0_{(1)}$ & $0_{(1)}$ & $0_{(1)}$ & $0_{(1)}$ & $0_{(1)}$ & $0_{(1)}$ & $0_{(1)}$ & $0_{(1)}$ & $y^2_{(1)}$ & $0_{(1)}$ & $0_{(1)}$ & $0_{(1)}$ & $0_{(1)}$ & $0_{(1)}$ & $0_{(1)}$ & \new $0_{(2)}$ & $\times_{(1)}$ & $(1)$ & $(1)$ & $(1)$ & $\times_{(1)}$ & $\times_{(1)}$ & $(1)$ & $(1)$ & $0_{(1)}$ & $0_{(1)}$ & $y^2_{(1)}$ & \new $\times_{(3)}$ \\
        $\psi^2 \phi^5$ & $(6,8)$ & $\times_{(1)}$ & $(1)$ & $(1)$ & $(1)$ & $\times_{(1)}$ & $\times_{(1)}$ & $(1)$ & $\times_{(1)}$ & $(1)$ & $(1)$ & $(1)$ & $(1)$ & $\times_{(1)}$ & $\times_{(1)}$ & $(1)$ & $(1)$ & $(1)$ & $(1)$ & $0_{(1)}$ & $(1)$ & $\times_{(1)}$ & $\times_{(1)}$ & $\times_{(1)}$ & $0_{(1)}$ & $0_{(1)}$ & $0_{(1)}$ & $(1)$ & $(1)$ & $(1)$ & $\times_{(1)}$ & $(1)$ & $\times_{(1)}$ & $0_{(1)}$ & $0_{(1)}$ & $0_{(1)}$ & $y^2_{(1)}$ & $\times_{(1)}$ & $(1)$ & $(1)$ & $(1)$ & $\times_{(1)}$ & $\times_{(1)}$ & $(1)$ & $(1)$ & $(1)$ & $(1)$ & $(1)$ & $(2)$ \\ \hline
        $\bar F^4$ & $(8,0)$ & \ccol$0_{(1)}$ & \ccol$0_{(1)}$ & \ccol$0_{(1)}$ & \ccol$0_{(1)}$ & \ccol$0_{(1)}$ & \new $0_{(2)}$ & \new $0_{(2)}$ & \new $0_{(2)}$ & \new $0_{(2)}$ & \new $0_{(2)}$ & \new $0_{(2)}$ & \new $0_{(2)}$ & \new $0_{(2)}$ & \new $0_{(2)}$ & \new $0_{(3)}$ & \new $0_{(3)}$ & \new $0_{(3)}$ & \new $0_{(4)}$ & $(1)$ & $0_{(1)}$ & $0_{(1)}$ & $0_{(1)}$ & $0_{(1)}$ & $(2)$ & $(2)$ & \new $\times_{(2)}$ & \new $0_{(2)}$ & \new $0_{(2)}$ & \new $0_{(2)}$ & \new $0_{(2)}$ & \new $0_{(2)}$ & \new $0_{(2)}$ & $(3)$ & \new $\times_{(3)}$ & \new $\times_{(3)}$ & \new $\times_{(4)}$ & $0_{(1)}$ & $0_{(1)}$ & $0_{(1)}$ & $0_{(1)}$ & $0_{(1)}$ & $0_{(1)}$ & $0_{(1)}$ & $0_{(1)}$ & \new $0_{(3)}$ & \new $0_{(3)}$ & \new $0_{(3)}$ & \new $\times_{(5)}$ \\
        $D^2 \phi^2 \bar F^2$ & $(6,2)$ & \ccol$0_{(1)}$ & $0_{(1)}$ & $0_{(1)}$ & $0_{(1)}$ & $0_{(1)}$ & \new $0_{(2)}$ & \new $0_{(2)}$ & \new $0_{(2)}$ & \new $0_{(2)}$ & \new $0_{(2)}$ & \new $0_{(2)}$ & \new $0_{(2)}$ & \new $0_{(2)}$ & \new $0_{(2)}$ & \new $0_{(3)}$ & \new $0_{(3)}$ & \new $\bar y^2_{(3)}$ & $(4)$ & $0_{(1)}$ & $(1)$ & $(1)$ & $\times_{(1)}$ & $(1)$ & $(2)$ & $(2)$ & \new $\times_{(2)}$ & $(2)$ & $(2)$ & $(2)$ & \new $\times_{(2)}$ & $(2)$ & \new $\times_{(2)}$ & $(3)$ & $(3)$ & $(3)$ & $(4)$ & $0_{(1)}$ & $0_{(1)}$ & $0_{(1)}$ & $0_{(1)}$ & $0_{(1)}$ & $0_{(1)}$ & $0_{(1)}$ & $0_{(1)}$ & $(3)$ & $(3)$ & $(3)$ & $(5)$ \\
        $D^2 \bar\psi^2 \phi \bar F$ & $(6,2)$ & \ccol$0_{(1)}$ & $0_{(1)}$ & $0_{(1)}$ & $0_{(1)}$ & $0_{(1)}$ & \new $0_{(2)}$ & \new $0_{(2)}$ & \new $0_{(2)}$ & \new $0_{(2)}$ & \new $0_{(2)}$ & \new $0_{(2)}$ & \new $\bar y^2_{(2)}$ & \new $0_{(2)}$ & \new $0_{(2)}$ & \new $0_{(3)}$ & \new $0_{(3)}$ & \new $0_{(3)}$ & \new $\times_{(4)}$ & $0_{(1)}$ & $(1)$ & $(1)$ & $(1)$ & $(1)$ & $(2)$ & $(2)$ & $(2)$ & $(2)$ & $(2)$ & $(2)$ & $(2)$ & $(2)$ & $(2)$ & $(3)$ & $(3)$ & $(3)$ & $(4)$ & $0_{(1)}$ & $0_{(1)}$ & $0_{(1)}$ & $\bar y^2_{(1)}$ & $0_{(1)}$ & $0_{(1)}$ & $0_{(1)}$ & $0_{(1)}$ & \new $\times_{(3)}$ & $(3)$ & $(3)$ & \new $\times_{(5)}$ \\
        $D^2 \bar\psi^4$ & $(6,2)$ & \ccol$0_{(1)}$ & $0_{(1)}$ & $0_{(1)}$ & $0_{(1)}$ & $0_{(1)}$ & \new $0_{(2)}$ & \new $0_{(2)}$ & \new $0_{(2)}$ & \new $0_{(2)}$ & \new $0_{(2)}$ & \new $0_{(2)}$ & \new $0_{(2)}$ & \new $0_{(2)}$ & \new $\bar y^2_{(2)}$ & \new $0_{(3)}$ & \new $0_{(3)}$ & \new $0_{(3)}$ & \new $\times_{(4)}$ & $0_{(1)}$ & $\times_{(1)}$ & $(1)$ & $(1)$ & $\times_{(1)}$ & \new $\times_{(2)}$ & $(2)$ & $(2)$ & \new $\times_{(2)}$ & $(2)$ & \new $\times_{(2)}$ & $(2)$ & \new $\times_{(2)}$ & $(2)$ & \new $\times_{(3)}$ & $(3)$ & $(3)$ & $(4)$ & $0_{(1)}$ & $0_{(1)}$ & $0_{(1)}$ & $0_{(1)}$ & $0_{(1)}$ & $0_{(1)}$ & $0_{(1)}$ & $\bar y^2_{(1)}$ & \new $\times_{(3)}$ & \new $\times_{(3)}$ & $(3)$ & \new $\times_{(5)}$ \\
        $D \bar\psi \psi \bar F^2$ & $(6,2)$ & \ccol$0_{(1)}$ & $0_{(1)}$ & $0_{(1)}$ & $0_{(1)}$ & $0_{(1)}$ & \new $0_{(2)}$ & \new $0_{(2)}$ & \new $0_{(2)}$ & \new $0_{(2)}$ & \new $0_{(2)}$ & \new $0_{(2)}$ & \new $0_{(2)}$ & \new $0_{(2)}$ & \new $0_{(2)}$ & \new $0_{(3)}$ & \new $0_{(3)}$ & \new $0_{(3)}$ & \new $\times_{(4)}$ & $0_{(1)}$ & $(1)$ & $(1)$ & $\times_{(1)}$ & $(1)$ & $(2)$ & $(2)$ & $(2)$ & \new $\times_{(2)}$ & \new $\times_{(2)}$ & $(2)$ & $(2)$ & $(2)$ & $(2)$ & $(3)$ & $(3)$ & \new $\times_{(3)}$ & \new $\times_{(4)}$ & $0_{(1)}$ & $0_{(1)}$ & $0_{(1)}$ & $0_{(1)}$ & $0_{(1)}$ & $0_{(1)}$ & $0_{(1)}$ & $0_{(1)}$ & \new $\times_{(3)}$ & $(3)$ & $(3)$ & \new $\times_{(5)}$ \\
        $\phi^2 \bar F^3$ & $(8,2)$ & \ccol$0_{(1)}$ & $0_{(1)}$ & $0_{(1)}$ & $0_{(1)}$ & $0_{(1)}$ & \ccol$0_{(1)}$ & \ccol$0_{(1)}$ & \ccol$0_{(1)}$ & $0_{(1)}$ & $0_{(1)}$ & $0_{(1)}$ & $0_{(1)}$ & $0_{(1)}$ & $0_{(1)}$ & \new $0_{(2)}$ & \new $0_{(2)}$ & \new $0_{(2)}$ & \new $0_{(3)}$ & $(1)$ & $(1)$ & $(1)$ & $\times_{(1)}$ & $(1)$ & $(1)$ & $(1)$ & $\times_{(1)}$ & $0_{(1)}$ & $0_{(1)}$ & $0_{(1)}$ & $0_{(1)}$ & $0_{(1)}$ & $0_{(1)}$ & $(2)$ & $(2)$ & \new $\times_{(2)}$ & \new $\times_{(3)}$ & $0_{(1)}$ & $0_{(1)}$ & $0_{(1)}$ & $0_{(1)}$ & $0_{(1)}$ & $0_{(1)}$ & $0_{(1)}$ & $0_{(1)}$ & \new $0_{(2)}$ & \new $0_{(2)}$ & \new $0_{(2)}$ & \new $\times_{(4)}$ \\
        $\bar\psi^2 \phi \bar F^2$ & $(8,2)$ & \ccol$0_{(1)}$ & $0_{(1)}$ & $0_{(1)}$ & $0_{(1)}$ & $0_{(1)}$ & \ccol$0_{(1)}$ & \ccol$0_{(1)}$ & \ccol$0_{(1)}$ & $0_{(1)}$ & $0_{(1)}$ & $0_{(1)}$ & $0_{(1)}$ & $0_{(1)}$ & $0_{(1)}$ & \new $0_{(2)}$ & \new $0_{(2)}$ & \new $0_{(2)}$ & \new $0_{(3)}$ & $(1)$ & $(1)$ & $(1)$ & $(1)$ & $(1)$ & $(1)$ & $(1)$ & $(1)$ & $0_{(1)}$ & $0_{(1)}$ & $0_{(1)}$ & $0_{(1)}$ & $\bar y^2_{(1)}$ & $0_{(1)}$ & $(2)$ & $(2)$ & $(2)$ & $(3)$ & $0_{(1)}$ & $0_{(1)}$ & $0_{(1)}$ & $\bar y^2_{(1)}$ & $0_{(1)}$ & $0_{(1)}$ & $0_{(1)}$ & $0_{(1)}$ & \new $0_{(2)}$ & \new $0_{(2)}$ & \new $0_{(2)}$ & \new $\times_{(4)}$ \\
        $\bar\psi^4 \bar F$ & $(8,2)$ & \ccol$0_{(1)}$ & $0_{(1)}$ & $0_{(1)}$ & $0_{(1)}$ & $0_{(1)}$ & \ccol$0_{(1)}$ & \ccol$0_{(1)}$ & \ccol$0_{(1)}$ & $0_{(1)}$ & $0_{(1)}$ & $0_{(1)}$ & $0_{(1)}$ & $0_{(1)}$ & $0_{(1)}$ & \new $0_{(2)}$ & \new $0_{(2)}$ & \new $0_{(2)}$ & \new $0_{(3)}$ & $\times_{(1)}$ & $\times_{(1)}$ & $(1)$ & $(1)$ & $(1)$ & $\times_{(1)}$ & $(1)$ & $(1)$ & $0_{(1)}$ & $0_{(1)}$ & $0_{(1)}$ & $0_{(1)}$ & $0_{(1)}$ & $\bar y^2_{(1)}$ & \new $\times_{(2)}$ & $(2)$ & $(2)$ & \new $\times_{(3)}$ & $0_{(1)}$ & $0_{(1)}$ & $0_{(1)}$ & $0_{(1)}$ & $0_{(1)}$ & $\bar y^2_{(1)}$ & $0_{(1)}$ & $\bar y^2_{(1)}$ & \new $0_{(2)}$ & \new $0_{(2)}$ & \new $0_{(2)}$ & \new $\times_{(4)}$ \\
        $D^2 \phi^4 \bar F$ & $(6,4)$ & $0_{(1)}$ & $0_{(1)}$ & $0_{(1)}$ & $0_{(1)}$ & $0_{(1)}$ & $0_{(1)}$ & $0_{(1)}$ & $0_{(1)}$ & $0_{(1)}$ & $0_{(1)}$ & $0_{(1)}$ & $0_{(1)}$ & $0_{(1)}$ & $0_{(1)}$ & \new $0_{(2)}$ & \new $0_{(2)}$ & \new $0_{(2)}$ & $(3)$ & $0_{(1)}$ & $(1)$ & $(1)$ & $\times_{(1)}$ & $\times_{(1)}$ & $0_{(1)}$ & $0_{(1)}$ & $0_{(1)}$ & $(1)$ & $(1)$ & $(1)$ & $\times_{(1)}$ & $\times_{(1)}$ & $\times_{(1)}$ & $(2)$ & $(2)$ & \new $\times_{(2)}$ & $(3)$ & $\times_{(1)}$ & $(1)$ & $(1)$ & $(1)$ & $\times_{(1)}$ & $\times_{(1)}$ & $(1)$ & $\times_{(1)}$ & $(2)$ & $(2)$ & \new $\times_{(2)}$ & $(4)$ \\
        $D^2 \bar\psi^2 \phi^3$ & $(6,4)$ & $0_{(1)}$ & $0_{(1)}$ & $0_{(1)}$ & $0_{(1)}$ & $0_{(1)}$ & $0_{(1)}$ & $0_{(1)}$ & $0_{(1)}$ & $0_{(1)}$ & $\bar y^2_{(1)}$ & $0_{(1)}$ & $0_{(1)}$ & $0_{(1)}$ & $0_{(1)}$ & \new $0_{(2)}$ & \new $\bar y^2_{(2)}$ & \new $0_{(2)}$ & $(3)$ & $0_{(1)}$ & $(1)$ & $(1)$ & $(1)$ & $\times_{(1)}$ & $0_{(1)}$ & $0_{(1)}$ & $0_{(1)}$ & $(1)$ & $(1)$ & $(1)$ & $(1)$ & $\times_{(1)}$ & $\times_{(1)}$ & $(2)$ & $(2)$ & $(2)$ & $(3)$ & $\times_{(1)}$ & $(1)$ & $(1)$ & $\times_{(1)}$ & $(1)$ & $\times_{(1)}$ & $(1)$ & $(1)$ & $(2)$ & $(2)$ & $(2)$ & $(4)$ \\
        $D \bar\psi \psi \phi^2 \bar F$ & $(6,4)$ & $0_{(1)}$ & $0_{(1)}$ & $0_{(1)}$ & $0_{(1)}$ & $0_{(1)}$ & $0_{(1)}$ & $0_{(1)}$ & $0_{(1)}$ & $0_{(1)}$ & $0_{(1)}$ & $0_{(1)}$ & $0_{(1)}$ & $0_{(1)}$ & $0_{(1)}$ & \new $0_{(2)}$ & \new $0_{(2)}$ & \new $\bar y^2_{(2)}$ & $(3)$ & $0_{(1)}$ & $(1)$ & $(1)$ & $\times_{(1)}$ & $(1)$ & $0_{(1)}$ & $0_{(1)}$ & $0_{(1)}$ & $(1)$ & $(1)$ & $(1)$ & $(1)$ & $(1)$ & $(1)$ & $(2)$ & $(2)$ & $(2)$ & $(3)$ & $\times_{(1)}$ & $(1)$ & $(1)$ & $(1)$ & $(1)$ & $(1)$ & $(1)$ & $(1)$ & \new $\times_{(2)}$ & $(2)$ & $(2)$ & \new $\times_{(4)}$ \\
        $D \bar\psi^3 \psi \phi$ & $(6,4)$ & $0_{(1)}$ & $0_{(1)}$ & $\bar y^2_{(1)}$ & $0_{(1)}$ & $0_{(1)}$ & $0_{(1)}$ & $0_{(1)}$ & $0_{(1)}$ & $0_{(1)}$ & $0_{(1)}$ & $0_{(1)}$ & $\bar y^2_{(1)}$ & $0_{(1)}$ & $0_{(1)}$ & \new $0_{(2)}$ & \new $0_{(2)}$ & \new $0_{(2)}$ & \new $\times_{(3)}$ & $0_{(1)}$ & $\times_{(1)}$ & $(1)$ & $(1)$ & $(1)$ & $0_{(1)}$ & $0_{(1)}$ & $0_{(1)}$ & $\times_{(1)}$ & $(1)$ & $(1)$ & $(1)$ & $\times_{(1)}$ & $(1)$ & \new $\times_{(2)}$ & $(2)$ & $(2)$ & $(3)$ & $\times_{(1)}$ & $\times_{(1)}$ & $\times_{(1)}$ & $(1)$ & $(1)$ & $(1)$ & $(1)$ & $(1)$ & \new $\times_{(2)}$ & $(2)$ & $(2)$ & \new $\times_{(4)}$ \\
        $\psi^2 \phi \bar F^2$ & $(6,4)$ & $0_{(1)}$ & $0_{(1)}$ & $0_{(1)}$ & $0_{(1)}$ & $0_{(1)}$ & $0_{(1)}$ & $0_{(1)}$ & $0_{(1)}$ & $0_{(1)}$ & $0_{(1)}$ & $0_{(1)}$ & $0_{(1)}$ & $0_{(1)}$ & $0_{(1)}$ & \new $0_{(2)}$ & \new $0_{(2)}$ & \new $0_{(2)}$ & $(3)$ & $0_{(1)}$ & $(1)$ & $(1)$ & $\times_{(1)}$ & $(1)$ & $0_{(1)}$ & $y^2_{(1)}$ & $0_{(1)}$ & $\times_{(1)}$ & $\times_{(1)}$ & $(1)$ & $\times_{(1)}$ & $(1)$ & $(1)$ & $(2)$ & \new $\times_{(2)}$ & \new $\times_{(2)}$ & \new $\times_{(3)}$ & $(1)$ & $(1)$ & $\times_{(1)}$ & $(1)$ & $\times_{(1)}$ & $(1)$ & $(1)$ & $(1)$ & \new $\times_{(2)}$ & \new $\times_{(2)}$ & $(2)$ & \new $\times_{(4)}$ \\
        $\bar\psi^2 \psi^2 \bar F$ & $(6,4)$ & $0_{(1)}$ & $0_{(1)}$ & $0_{(1)}$ & $\bar y^2_{(1)}$ & $0_{(1)}$ & $0_{(1)}$ & $0_{(1)}$ & $0_{(1)}$ & $0_{(1)}$ & $0_{(1)}$ & $0_{(1)}$ & $0_{(1)}$ & $0_{(1)}$ & $0_{(1)}$ & \new $0_{(2)}$ & \new $0_{(2)}$ & \new $0_{(2)}$ & \new $\times_{(3)}$ & $0_{(1)}$ & $\times_{(1)}$ & $(1)$ & $(1)$ & $(1)$ & $0_{(1)}$ & $0_{(1)}$ & $y^2_{(1)}$ & $\times_{(1)}$ & $\times_{(1)}$ & $(1)$ & $(1)$ & $(1)$ & $(1)$ & \new $\times_{(2)}$ & $(2)$ & \new $\times_{(2)}$ & \new $\times_{(3)}$ & $\times_{(1)}$ & $\times_{(1)}$ & $\times_{(1)}$ & $(1)$ & $(1)$ & $(1)$ & $(1)$ & $(1)$ & \new $\times_{(2)}$ & \new $\times_{(2)}$ & $(2)$ & \new $\times_{(4)}$ \\
        $\phi^4 \bar F^2$ & $(8,4)$ & $0_{(1)}$ & $0_{(1)}$ & $0_{(1)}$ & $0_{(1)}$ & $0_{(1)}$ & $0_{(1)}$ & $0_{(1)}$ & $0_{(1)}$ & $0_{(1)}$ & $0_{(1)}$ & $0_{(1)}$ & $0_{(1)}$ & $0_{(1)}$ & $0_{(1)}$ & $0_{(1)}$ & $0_{(1)}$ & $0_{(1)}$ & \new $0_{(2)}$ & $(1)$ & $(1)$ & $(1)$ & $\times_{(1)}$ & $(1)$ & $(1)$ & $(1)$ & $\times_{(1)}$ & $(1)$ & $(1)$ & $(1)$ & $\times_{(1)}$ & $(1)$ & $\times_{(1)}$ & $(1)$ & $(1)$ & $\times_{(1)}$ & $(2)$ & $(1)$ & $(1)$ & $(1)$ & $(1)$ & $\times_{(1)}$ & $\times_{(1)}$ & $(1)$ & $\times_{(1)}$ & $0_{(1)}$ & $0_{(1)}$ & $0_{(1)}$ & $(3)$ \\
        $\bar\psi^2 \phi^3 \bar F$ & $(8,4)$ & $0_{(1)}$ & $0_{(1)}$ & $0_{(1)}$ & $0_{(1)}$ & $0_{(1)}$ & $0_{(1)}$ & $0_{(1)}$ & $0_{(1)}$ & $0_{(1)}$ & $\bar y^2_{(1)}$ & $0_{(1)}$ & $0_{(1)}$ & $0_{(1)}$ & $0_{(1)}$ & $0_{(1)}$ & $0_{(1)}$ & $0_{(1)}$ & \new $0_{(2)}$ & $\times_{(1)}$ & $(1)$ & $(1)$ & $(1)$ & $(1)$ & $(1)$ & $(1)$ & $(1)$ & $(1)$ & $(1)$ & $(1)$ & $(1)$ & $\times_{(1)}$ & $(1)$ & $(1)$ & $(1)$ & $(1)$ & $(2)$ & $\times_{(1)}$ & $(1)$ & $(1)$ & $(1)$ & $(1)$ & $(1)$ & $(1)$ & $(1)$ & $0_{(1)}$ & $0_{(1)}$ & $0_{(1)}$ & \new $\times_{(3)}$ \\
        $\bar\psi^4 \phi^2$ & $(8,4)$ & $0_{(1)}$ & $\bar y^2_{(1)}$ & $0_{(1)}$ & $0_{(1)}$ & $0_{(1)}$ & $0_{(1)}$ & $0_{(1)}$ & $0_{(1)}$ & $0_{(1)}$ & $0_{(1)}$ & $\bar y^2_{(1)}$ & $0_{(1)}$ & $0_{(1)}$ & $0_{(1)}$ & $0_{(1)}$ & $0_{(1)}$ & $0_{(1)}$ & \new $0_{(2)}$ & $\times_{(1)}$ & $(1)$ & $(1)$ & $(1)$ & $\times_{(1)}$ & $\times_{(1)}$ & $(1)$ & $(1)$ & $\times_{(1)}$ & $(1)$ & $(1)$ & $(1)$ & $\times_{(1)}$ & $\times_{(1)}$ & $\times_{(1)}$ & $(1)$ & $(1)$ & $(2)$ & $\times_{(1)}$ & $(1)$ & $(1)$ & $\times_{(1)}$ & $(1)$ & $\times_{(1)}$ & $(1)$ & $(1)$ & $0_{(1)}$ & $0_{(1)}$ & $\bar y^2_{(1)}$ & \new $\times_{(3)}$ \\
        $\bar\psi^2 \phi^5$ & $(8,6)$ & $0_{(1)}$ & $(1)$ & $\times_{(1)}$ & $\times_{(1)}$ & $\times_{(1)}$ & $0_{(1)}$ & $0_{(1)}$ & $0_{(1)}$ & $(1)$ & $(1)$ & $(1)$ & $\times_{(1)}$ & $(1)$ & $\times_{(1)}$ & $0_{(1)}$ & $0_{(1)}$ & $0_{(1)}$ & $\bar y^2_{(1)}$ & $\times_{(1)}$ & $(1)$ & $(1)$ & $(1)$ & $\times_{(1)}$ & $\times_{(1)}$ & $(1)$ & $\times_{(1)}$ & $(1)$ & $(1)$ & $(1)$ & $(1)$ & $\times_{(1)}$ & $\times_{(1)}$ & $(1)$ & $(1)$ & $(1)$ & $(1)$ & $\times_{(1)}$ & $(1)$ & $(1)$ & $\times_{(1)}$ & $(1)$ & $\times_{(1)}$ & $(1)$ & $(1)$ & $(1)$ & $(1)$ & $(1)$ & $(2)$ \\ \hline
        $F^2 \bar F^2$ & $(4,4)$ & $0_{(1)}$ & $0_{(1)}$ & $0_{(1)}$ & $0_{(1)}$ & $0_{(1)}$ & \new $0_{(2)}$ & \new $0_{(2)}$ & \new $0_{(2)}$ & \new $\times_{(2)}$ & \new $\times_{(2)}$ & \new $\times_{(2)}$ & \new $\times_{(2)}$ & $(2)$ & \new $\times_{(2)}$ & $(3)$ & \new $\times_{(3)}$ & \new $\times_{(3)}$ & \new $\times_{(4)}$ & $0_{(1)}$ & $0_{(1)}$ & $0_{(1)}$ & $0_{(1)}$ & $0_{(1)}$ & \new $0_{(2)}$ & \new $0_{(2)}$ & \new $0_{(2)}$ & \new $\times_{(2)}$ & \new $\times_{(2)}$ & \new $\times_{(2)}$ & \new $\times_{(2)}$ & $(2)$ & \new $\times_{(2)}$ & $(3)$ & \new $\times_{(3)}$ & \new $\times_{(3)}$ & \new $\times_{(4)}$ & $(1)$ & $(1)$ & $\times_{(1)}$ & $\times_{(1)}$ & $\times_{(1)}$ & $(1)$ & $\times_{(1)}$ & $\times_{(1)}$ & \new $\times_{(3)}$ & \new $\times_{(3)}$ & \new $\times_{(3)}$ & \new $\times_{(5)}$ \\
        $D^2 \phi^2 F \bar F$ & $(4,4)$ & $0_{(1)}$ & $0_{(1)}$ & $0_{(1)}$ & $0_{(1)}$ & $0_{(1)}$ & \new $0_{(2)}$ & \new $0_{(2)}$ & \new $0_{(2)}$ & $(2)$ & $(2)$ & $(2)$ & \new $\times_{(2)}$ & $(2)$ & \new $\times_{(2)}$ & $(3)$ & $(3)$ & $(3)$ & $(4)$ & $0_{(1)}$ & $0_{(1)}$ & $0_{(1)}$ & $0_{(1)}$ & $0_{(1)}$ & \new $0_{(2)}$ & \new $0_{(2)}$ & \new $0_{(2)}$ & $(2)$ & $(2)$ & $(2)$ & \new $\times_{(2)}$ & $(2)$ & \new $\times_{(2)}$ & $(3)$ & $(3)$ & $(3)$ & $(4)$ & $(1)$ & $(1)$ & $(1)$ & $(1)$ & $(1)$ & $(1)$ & $(1)$ & $\times_{(1)}$ & $(3)$ & $(3)$ & $(3)$ & $(5)$ \\
        $D^4 \phi^4$ & $(4,4)$ & $0_{(1)}$ & $0_{(1)}$ & $0_{(1)}$ & $0_{(1)}$ & $0_{(1)}$ & \new $0_{(2)}$ & \new $0_{(2)}$ & \new $0_{(2)}$ & $(2)$ & $(2)$ & $(2)$ & \new $\times_{(2)}$ & \new $\times_{(2)}$ & \new $\times_{(2)}$ & $(3)$ & $(3)$ & $(3)$ & $(4)$ & $0_{(1)}$ & $0_{(1)}$ & $0_{(1)}$ & $0_{(1)}$ & $0_{(1)}$ & \new $0_{(2)}$ & \new $0_{(2)}$ & \new $0_{(2)}$ & $(2)$ & $(2)$ & $(2)$ & \new $\times_{(2)}$ & \new $\times_{(2)}$ & \new $\times_{(2)}$ & $(3)$ & $(3)$ & $(3)$ & $(4)$ & $\times_{(1)}$ & $(1)$ & $(1)$ & $\times_{(1)}$ & $\times_{(1)}$ & $\times_{(1)}$ & $(1)$ & $\times_{(1)}$ & $(3)$ & $(3)$ & $(3)$ & $(5)$ \\
        $D^2 \psi^2 \phi \bar F$ & $(4,4)$ & $0_{(1)}$ & $0_{(1)}$ & $0_{(1)}$ & $0_{(1)}$ & $0_{(1)}$ & \new $0_{(2)}$ & \new $0_{(2)}$ & \new $0_{(2)}$ & \new $\times_{(2)}$ & $(2)$ & $(2)$ & $(2)$ & \new $\times_{(2)}$ & $(2)$ & \new $\times_{(3)}$ & $(3)$ & $(3)$ & $(4)$ & $0_{(1)}$ & $0_{(1)}$ & $y^2_{(1)}$ & $0_{(1)}$ & $0_{(1)}$ & \new $0_{(2)}$ & \new $y^2_{(2)}$ & \new $0_{(2)}$ & $(2)$ & \new $\times_{(2)}$ & $(2)$ & $(2)$ & $(2)$ & $(2)$ & $(3)$ & $(3)$ & \new $\times_{(3)}$ & \new $\times_{(4)}$ & $\times_{(1)}$ & $(1)$ & $\times_{(1)}$ & $(1)$ & $\times_{(1)}$ & $(1)$ & $(1)$ & $(1)$ & \new $\times_{(3)}$ & $(3)$ & $(3)$ & \new $\times_{(5)}$ \\
        $D^2 \bar\psi^2 \phi F$ & $(4,4)$ & $0_{(1)}$ & $0_{(1)}$ & $\bar y^2_{(1)}$ & $0_{(1)}$ & $0_{(1)}$ & \new $0_{(2)}$ & \new $\bar y^2_{(2)}$ & \new $0_{(2)}$ & $(2)$ & \new $\times_{(2)}$ & $(2)$ & $(2)$ & $(2)$ & $(2)$ & $(3)$ & $(3)$ & \new $\times_{(3)}$ & \new $\times_{(4)}$ & $0_{(1)}$ & $0_{(1)}$ & $0_{(1)}$ & $0_{(1)}$ & $0_{(1)}$ & \new $0_{(2)}$ & \new $0_{(2)}$ & \new $0_{(2)}$ & \new $\times_{(2)}$ & $(2)$ & $(2)$ & $(2)$ & \new $\times_{(2)}$ & $(2)$ & \new $\times_{(3)}$ & $(3)$ & $(3)$ & $(4)$ & $\times_{(1)}$ & $(1)$ & $\times_{(1)}$ & $\times_{(1)}$ & $(1)$ & $(1)$ & $(1)$ & $(1)$ & \new $\times_{(3)}$ & $(3)$ & $(3)$ & \new $\times_{(5)}$ \\
        $D \bar\psi \psi F \bar F$ & $(4,4)$ & $0_{(1)}$ & $0_{(1)}$ & $0_{(1)}$ & $0_{(1)}$ & $0_{(1)}$ & \new $0_{(2)}$ & \new $0_{(2)}$ & \new $\bar y^2_{(2)}$ & \new $\times_{(2)}$ & \new $\times_{(2)}$ & $(2)$ & $(2)$ & $(2)$ & $(2)$ & \new $\times_{(3)}$ & $(3)$ & \new $\times_{(3)}$ & \new $\times_{(4)}$ & $0_{(1)}$ & $0_{(1)}$ & $0_{(1)}$ & $0_{(1)}$ & $0_{(1)}$ & \new $0_{(2)}$ & \new $0_{(2)}$ & \new $y^2_{(2)}$ & \new $\times_{(2)}$ & \new $\times_{(2)}$ & $(2)$ & $(2)$ & $(2)$ & $(2)$ & \new $\times_{(3)}$ & $(3)$ & \new $\times_{(3)}$ & \new $\times_{(4)}$ & $(1)$ & $(1)$ & $\times_{(1)}$ & $(1)$ & $(1)$ & $(1)$ & $(1)$ & $(1)$ & \new $\times_{(3)}$ & $(3)$ & $(3)$ & \new $\times_{(5)}$ \\
        $D^3 \bar\psi \psi \phi^2$ & $(4,4)$ & $0_{(1)}$ & $0_{(1)}$ & $0_{(1)}$ & $0_{(1)}$ & $0_{(1)}$ & \new $0_{(2)}$ & \new $0_{(2)}$ & \new $0_{(2)}$ & $(2)$ & $(2)$ & $(2)$ & $(2)$ & $(2)$ & $(2)$ & $(3)$ & $(3)$ & $(3)$ & $(4)$ & $0_{(1)}$ & $0_{(1)}$ & $0_{(1)}$ & $0_{(1)}$ & $0_{(1)}$ & \new $0_{(2)}$ & \new $0_{(2)}$ & \new $0_{(2)}$ & $(2)$ & $(2)$ & $(2)$ & $(2)$ & $(2)$ & $(2)$ & $(3)$ & $(3)$ & $(3)$ & $(4)$ & $\times_{(1)}$ & $(1)$ & $(1)$ & $(1)$ & $(1)$ & $(1)$ & $(1)$ & $(1)$ & $(3)$ & $(3)$ & $(3)$ & $(5)$ \\
        $D^2 \bar\psi^2 \psi^2$ & $(4,4)$ & $0_{(1)}$ & $0_{(1)}$ & $0_{(1)}$ & $\bar y^2_{(1)}$ & $0_{(1)}$ & \new $0_{(2)}$ & \new $0_{(2)}$ & \new $\bar y^2_{(2)}$ & \new $\times_{(2)}$ & $(2)$ & $(2)$ & $(2)$ & $(2)$ & $(2)$ & \new $\times_{(3)}$ & $(3)$ & $(3)$ & $(4)$ & $0_{(1)}$ & $0_{(1)}$ & $0_{(1)}$ & $y^2_{(1)}$ & $0_{(1)}$ & \new $0_{(2)}$ & \new $0_{(2)}$ & \new $y^2_{(2)}$ & \new $\times_{(2)}$ & $(2)$ & $(2)$ & $(2)$ & $(2)$ & $(2)$ & \new $\times_{(3)}$ & $(3)$ & $(3)$ & $(4)$ & $\times_{(1)}$ & $\times_{(1)}$ & $\times_{(1)}$ & $(1)$ & $(1)$ & $(1)$ & $(1)$ & $(1)$ & \new $\times_{(3)}$ & $(3)$ & $(3)$ & \new $\times_{(5)}$ \\
        $D^2 \phi^6$ & $(6,6)$ & $0_{(1)}$ & $(1)$ & $\times_{(1)}$ & $\times_{(1)}$ & $\times_{(1)}$ & $0_{(1)}$ & $0_{(1)}$ & $0_{(1)}$ & $(1)$ & $(1)$ & $\times_{(1)}$ & $\times_{(1)}$ & $\times_{(1)}$ & $\times_{(1)}$ & $0_{(1)}$ & $0_{(1)}$ & $0_{(1)}$ & $(2)$ & $0_{(1)}$ & $(1)$ & $\times_{(1)}$ & $\times_{(1)}$ & $\times_{(1)}$ & $0_{(1)}$ & $0_{(1)}$ & $0_{(1)}$ & $(1)$ & $(1)$ & $\times_{(1)}$ & $\times_{(1)}$ & $\times_{(1)}$ & $\times_{(1)}$ & $0_{(1)}$ & $0_{(1)}$ & $0_{(1)}$ & $(2)$ & $\times_{(1)}$ & $(1)$ & $(1)$ & $\times_{(1)}$ & $\times_{(1)}$ & $\times_{(1)}$ & $(1)$ & $\times_{(1)}$ & $(1)$ & $(1)$ & $\times_{(1)}$ & $(3)$ \\
        $D \bar\psi \psi \phi^4$ & $(6,6)$ & $0_{(1)}$ & $(1)$ & $(1)$ & $\times_{(1)}$ & $(1)$ & $0_{(1)}$ & $0_{(1)}$ & $0_{(1)}$ & $(1)$ & $(1)$ & $(1)$ & $(1)$ & $\times_{(1)}$ & $\times_{(1)}$ & $0_{(1)}$ & $0_{(1)}$ & $0_{(1)}$ & $(2)$ & $0_{(1)}$ & $(1)$ & $(1)$ & $\times_{(1)}$ & $(1)$ & $0_{(1)}$ & $0_{(1)}$ & $0_{(1)}$ & $(1)$ & $(1)$ & $(1)$ & $(1)$ & $\times_{(1)}$ & $\times_{(1)}$ & $0_{(1)}$ & $0_{(1)}$ & $0_{(1)}$ & $(2)$ & $\times_{(1)}$ & $(1)$ & $(1)$ & $(1)$ & $(1)$ & $(1)$ & $(1)$ & $(1)$ & $(1)$ & $(1)$ & $(1)$ & $(3)$ \\
        $\bar\psi^2 \psi^2 \phi^2$ & $(6,6)$ & $0_{(1)}$ & $(1)$ & $(1)$ & $(1)$ & $(1)$ & $0_{(1)}$ & $0_{(1)}$ & $0_{(1)}$ & $\times_{(1)}$ & $(1)$ & $(1)$ & $(1)$ & $(1)$ & $(1)$ & $0_{(1)}$ & $0_{(1)}$ & $\bar y^2_{(1)}$ & $(2)$ & $0_{(1)}$ & $(1)$ & $(1)$ & $(1)$ & $(1)$ & $0_{(1)}$ & $0_{(1)}$ & $0_{(1)}$ & $\times_{(1)}$ & $(1)$ & $(1)$ & $(1)$ & $(1)$ & $(1)$ & $0_{(1)}$ & $0_{(1)}$ & $y^2_{(1)}$ & $(2)$ & $\times_{(1)}$ & $(1)$ & $(1)$ & $(1)$ & $(1)$ & $(1)$ & $(1)$ & $(1)$ & $\times_{(1)}$ & $(1)$ & $(1)$ & \new $\times_{(3)}$ \\
        $\phi^8$ & $(8,8)$ & $\times_{(1)}$ & $(1)$ & $\times_{(1)}$ & $\times_{(1)}$ & $\times_{(1)}$ & $\times_{(1)}$ & $\times_{(1)}$ & $\times_{(1)}$ & $(1)$ & $(1)$ & $\times_{(1)}$ & $\times_{(1)}$ & $\times_{(1)}$ & $\times_{(1)}$ & $(1)$ & $\times_{(1)}$ & $\times_{(1)}$ & $(1)$ & $\times_{(1)}$ & $(1)$ & $\times_{(1)}$ & $\times_{(1)}$ & $\times_{(1)}$ & $\times_{(1)}$ & $\times_{(1)}$ & $\times_{(1)}$ & $(1)$ & $(1)$ & $\times_{(1)}$ & $\times_{(1)}$ & $\times_{(1)}$ & $\times_{(1)}$ & $(1)$ & $\times_{(1)}$ & $\times_{(1)}$ & $(1)$ & $\times_{(1)}$ & $(1)$ & $(1)$ & $\times_{(1)}$ & $\times_{(1)}$ & $\times_{(1)}$ & $(1)$ & $\times_{(1)}$ & $(1)$ & $(1)$ & $\times_{(1)}$ & $(1)$ \\
    \end{NiceTabular}
    }
    \caption{Structure of the anomalous-dimension matrix for dimension-eight operators. The notation is described in Table \ref{tab:dim5}.}
    \label{tab:dim8}
\end{table*}
\end{turnpage}

\pdfbookmark[1]{Discussion and outlook}{Discussion and outlook}
\paragraph{\textbf{Discussion and outlook.}}

We have established new non-renormalization theorems at higher loops, showing that helicity remains an active constraint on operator mixings beyond one loop. Besides uncovering broad classes of previously unknown zeros, the theorems already provide sharp checks on recent explicit calculations. 
In particular, the scheme-independent zeros discovered here provide an independent four-dimensional handle for validating the treatment of potentially ambiguous $\gamma_5$-odd traces; see, e.g., Ref.~\cite{Born:2026xkr}. 

Beyond its immediate use as a consistency check, the structure uncovered here opens several avenues for further study. 
For the helicity zeros in the scheme-independent region, the corresponding three-particle interference vanishes at the integrand level, suggesting phenomenological non-interference patterns worth exploring \cite{Azatov:2016sqh}. 
Moreover, since this region is controlled entirely by three-particle cuts of tree-level objects, decomposing these cuts into generalized partial waves \cite{Jiang:2020rwz,Shu:2021qlr,Bresciani:2025toe} may reveal additional scheme-independent two-loop zeros based on angular momentum \cite{Jiang:2020rwz}. 
Further extensions include non-linear mixing from multiple operator insertions and the scheme-dependent sector beyond two loops, where the generalization of \Eq{eq:dl2} remains to be established.
Finally, the modified-helicity framework of Ref.~\cite{Baratella:2021guc} places minimally coupled gravitational amplitudes under the same weight bounds used here, 
suggesting no obvious obstruction to extending the theorems to theories with spin-$2$ states. \smallskip

\pdfbookmark[1]{Acknowledgments}{Acknowledgments}
\paragraph{\textbf{Acknowledgments.}}
We thank Zvi Bern, Gauthier Durieux, Admir Greljo, Paride Paradisi, Julio Parra-Martinez, and Chia-Hsien Shen for useful discussions.
The project received funding from the Istituto Nazionale di Fisica Nucleare (INFN) Iniziativa Specifica APINE. N.S.\ is supported by the Italian MUR through the FIS 2 project FIS-2023-01577 (DD n.\ 23314 10-12-2024, CUP C53C24001460001), and by INFN through the Theoretical Astroparticle Physics (TAsP) project. We acknowledge
support from the COMETA COST Action CA22130. \smallskip

\pdfbookmark[1]{Appendix: Construction of the holomorphic scheme}{Appendix: Construction of the holomorphic scheme}
\paragraph{\textbf{Appendix: Construction of the holomorphic scheme.}}

In this appendix we prove the decomposition in Eq.~\eqref{eq:contact} and use it to establish Eqs.~\eqref{eq:dl2} and \eqref{eq:onedet}.

Under a finite renormalization
$\hat{\mathcal O}_j = \sum_i Z_{ij}\,\mathcal O_i$, with $Z_{ij} = \delta_{ij} + f^{(1)}_{ij} + \dots$, the two-loop anomalous-dimension matrix transforms as \cite{Bern:2020ikv}
\begin{equation}
\hat\gamma^{(2)}_{ij}
=\gamma^{(2)}_{ij}
-[f^{(1)},\gamma^{(1)}]_{ij}
-\beta^{(1)}_g\partial_g f^{(1)}_{ij}\,,
\label{eq:cov}
\end{equation}
where $[f^{(1)},\gamma^{(1)}]_{ij} = \sum_k (f^{(1)}_{ik}\gamma^{(1)}_{kj} - \gamma^{(1)}_{ik}f^{(1)}_{kj})$, while one-loop and IR anomalous dimensions are unaffected.

The derivation of \Eq{eq:dl2} proceeds in two steps: first, we bound the logarithmic and rational parts of the one-loop objects entering \Eq{eq:master}, and, second, we show that the only contributions reaching the region where $w_i < w_j - 4$ or $\wb_i < \wb_j - 4$ have exactly the form of the last two terms of \Eq{eq:cov}, so that a choice of $f^{(1)}$ can remove them.

The amplitude $\M^{(1)}$ entering cut (a) is bounded immediately.
Its logarithms are cut-constructible, hence built from tree amplitudes, and floor at $w,\wb = 4$ ($2$ with non-holomorphic Yukawa couplings), while its rational part is constrained only by the spin bound $w,\wb \ge 0$, saturated by the all-plus and -minus vector amplitudes.
Therefore, by \Eq{eq:sewing}, cut (a) floors at $w_i = w_j - 4$ and $\wb_i = \wb_j - 4$.

The form factor $F^{(1)}_j$, which contributes to cut (b) and the $\beta$-function and iteration terms of \Eq{eq:master}, requires more care.
Consider $F^{(1)}_j(X)$ on a configuration $X$ with $w(X) < w_j$ or $\wb(X) < \wb_j$.
Its logarithms are fixed by the two-particle discontinuities, each a product of trees: in the channel with a subset of legs $T$, 
the discontinuity in $s_T = (\sum_{a \in T} p_a)^2$ factorizes into a renormalizable tree amplitude of at least four legs ($w,\wb \ge 4$)
glued to a tree form factor of $\Op_j$ ($w \ge w_j$ and $\wb \ge \wb_j$).
By \Eq{eq:sewing} every discontinuity then requires $w(X) \ge w_j$ and $\wb(X) \ge \wb_j$---the one-loop selection rule of \Eq{eq:1lhsr} applied to the cut---so, below the bound, all discontinuities vanish and, with them, every logarithm and every UV and IR divergence \cite{Craig:2019wmo}.
The form factor is thus finite and purely rational for $w(X) < w_j$ or $\wb(X) < \wb_j$, or, in the presence of non-holomorphic Yukawa couplings, for $w(X) <w_j-2$ or $\wb(X) <\wb_j-2$.
Four-dimensional cuts do not determine this rational part \cite{Bern:1994zx}, 
but its kinematic poles can determine its structure, up to local terms.

In the region $w(X)<w_j-2$ or $\wb(X)<\wb_j-2$, the stronger statement is the decomposition in \Eq{eq:contact} of the main text.
Moreover, the coefficients $\kappa^{(1)}_{ij}$ depend on the treatment of the evanescent sector, i.e., the $\gamma_5$ and Levi-Civita prescriptions and the subtraction of evanescent operators \cite{Buras:1989xd,Dugan:1990df,Herrlich:1994kh}.
\Eq{eq:contact} states that all the singularities of the rational part of $F^{(1)}_j(X)$ are exactly those of tree structures of lower-weight operators with coefficients that are independent of 
multiplicity and 
kinematics, which is what a finite operator renormalization generates.

We prove \Eq{eq:contact} by induction on the number of legs $n$ of $X$.
Being purely rational in the region where $w(X) < w_j - 2$ or $\wb(X) < \wb_j - 2$, $F^{(1)}_j(X)$ can have at most kinematic poles. 
At the minimal multiplicity $n = \ell_j$, any such pole would involve a form factor with fewer than $\ell_j$ legs, which vanishes identically since, at tree level, it is forbidden by particle content and, at one loop, every candidate diagram contains a scaleless integral \cite{Bern:2019wie}.
$F^{(1)}_j(X)$ is therefore regular, i.e., a polynomial, and matching
it onto the independent contact structures of dimension-$d$ operators defines the constants
$\kappa^{(1)}_{ij}$ for $\ell_i = \ell_j$.
(For $\ell_i < \ell_j$ the matching occurs at the target multiplicity $\ell_i$, where
$F^{(1)}_j$ vanishes identically by the same leg counting, forcing $\kappa^{(1)}_{ij} = 0$ in this region.)
%
\begin{figure}[t]
    \centering
    \includegraphics[width=0.90\linewidth,page=2]{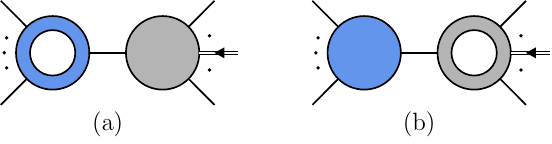}
    \caption{Factorization channels of the one-loop form factor $F^{(1)}_j$ with $w<w_j-2$ or $\wb < \wb_j - 2$. (a)~One-loop amplitude glued to a tree form factor, which is forbidden by the weight
	counting of \Eq{eq:sewing}. (b)~Tree amplitude glued to a lower-point one-loop form factor, which is
	reduced to contact terms by the induction hypothesis.}
	\label{fig:poles}
\end{figure}
%

For $n > \ell_j$, assume that \Eq{eq:contact} holds at all lower multiplicities and consider the subtracted object $\hat F^{(1)}_j(X) = F^{(1)}_j(X) - \sum_i \kappa^{(1)}_{ij} F^{(0)}_i(X)$, with the constants fixed at those multiplicities.
In any factorization channel of $X$, at least two legs move to the amplitude side, leaving the operator insertion on a configuration $Y$ with fewer than $n$ legs. Sewing the two corners across the single internal line gives $w(X) = w(\M) + w(Y) - 2$.
The residue splits by loop order, as illustrated in \Fig{fig:poles}: (a) a one-loop amplitude times a tree form factor, or (b) a tree amplitude times a lower-point one-loop form factor.\footnote{One-loop factorization allows for a third structure, not shown: a tree amplitude times a tree form
factor dressed by a universal factorization function \cite{Bern:1995ix}.
However, this term carries no helicity weight and therefore obeys the tree bound $w(X)\ge w_j$.}
The first split requires $w(X) = w(\M^{(1)}) + w(Y) - 2 \ge 0 + w_j - 2$ and is thus absent below the bound.
In the second split, $w(\M^{(0)}) \ge 2$ implies $w(Y) \le w(X) < w_j - 2$, so that the lower-point form factor is evaluated below the bound, where the inductive hypothesis applies. Since the tree structures $F^{(0)}_i(X)$ factorize in the same channel with the same tree amplitude corner, the residue of the subtracted object is $\M^{(0)} \hat F^{(1)}_j(Y)$, which vanishes.
Having no residue in any channel, $\hat F^{(1)}_j(X)$ is a new contact structure that is absorbed into $\kappa^{(1)}_{ij}$ for targets with $\ell_i = n$.
Subject to the complex-collinear assumption specified below, this closes the induction and establishes \Eq{eq:contact}.

One caveat concerns the two-particle channels.
Residues factorize rigorously on multi-particle channels, where the factorization kinematics are non-degenerate.
On the two-particle channel of a pair $(a,b)$, one-loop rational terms exhibit a richer singular behavior \cite{Bern:2005hs,Bern:2005ji,Bern:2005cq}: besides the pole in $s_{ab}$, they develop single and double poles in a single bracket, $\agl{a}{b}^{-1}$ and $\agl{a}{b}^{-2}$, at generic $\sqr{a}{b}$ (and conjugates).
The double poles have no tree-level counterparts, and the single-bracket single poles---absent at real momenta and known as unreal poles---have residues that are neither fixed by factorization nor controlled by a general theorem.
We assume that, on configurations below the bound, this singular behavior nevertheless retains the factorized form of \Fig{fig:poles}, with $a$ and $b$ merged into a single on-shell leg: an amplitude corner times a lower-point form factor of $\Op_j$, possibly dressed by functions carrying no little-group weight, which do not affect the counting.
Only the little-group weights of the residues are used: the corner is a three-point object, carrying $w,\wb \ge 2$ at tree level like every renormalizable tree amplitude and $w,\wb \ge 0$ at one loop, so the two splits above apply unchanged and the singular behavior vanishes below the bound.
At leading power in real collinear kinematics, this factorized form is a theorem for amplitudes \cite{Bern:1995ix,Kosower:1999xi}. However, for complex momenta, no general theorem exists, and our results for $\ell_i \ge \ell_j$ depend on this assumption.

We can now analyze what can reach the region where $w_i < w_j - 4$ or $\wb_i < \wb_j - 4$: cuts (a)
and (c) are absent since they floor at $w_j - 4$ and $w_j-2$, respectively (cut (c) floors at $w_j - 4$ as well with non-holomorphic Yukawa couplings), 
the logarithmic and non-contact rational parts (i.e., those not of the form of \Eq{eq:contact}) of
the $\beta$-function terms floor at $w_j - 2$, and the logarithmic and
non-contact rational parts of the iteration terms and cut (b) floor at $w_j - 4$.\footnote{The iteration term $\Delta\gamma^{(1)}_{kj}\Re F^{(1)}_k(X_i)$ requires $w_i \ge w_k - 2 \ge w_j - 4$, where the last inequality follows from the one-loop helicity selection rule, weakened by the presence of non-holomorphic Yukawa couplings, applied to $\Delta\gamma^{(1)}_{kj}$. The bounds apply equally to $\wb$.}
Thus, the only contributions that survive are
the contact terms of the one-loop form factors in cut (b)
and in the $\beta$-function and iteration terms, which can be interpreted as a finite renormalization effect.
Removing them by the choice $f^{(1)}_{ij} = - \kappa^{(1)}_{ij}$ defines a scheme
in which every term of the master formula of \Eq{eq:master} vanishes in this region, establishing
\Eq{eq:dl2}, which is robust against non-holomorphic Yukawa couplings.

Finally, substituting $f_{ij}^{(1)}=-\kappa_{ij}^{(1)}$ and $\hat\gamma^{(2)}_{ij}=0$ into \Eq{eq:cov} gives \Eq{eq:onedet} in the stated region.

\bibliographystyle{JHEP}
\bibliography{bibliography}

\end{document}